\documentclass[aps,pra,amsmath,amssymb,tightenlines,epsfig,floatfix,twocolumn,superscriptaddress, longbibliography]{revtex4-1}
\usepackage{graphicx}
\usepackage{dcolumn}	
\usepackage{bm}			
\usepackage{amsfonts}
\usepackage{xspace}
\usepackage{color}
\usepackage{epstopdf}
\usepackage{multirow}

\begin{document}

	\newcommand{\psihat}{\ensuremath{\hat{\psi}}\xspace}
	\newcommand{\psihatd}{\ensuremath{\hat{\psi}^{\dagger}}\xspace}
	\newcommand{\ahat}{\ensuremath{\hat{a}}\xspace}
	\newcommand{\Ham}{\ensuremath{\mathcal{H}}\xspace}
	\newcommand{\ahatd}{\ensuremath{\hat{a}^{\dagger}}\xspace}
	\newcommand{\bhat}{\ensuremath{\hat{b}}\xspace}
	\newcommand{\bhatd}{\ensuremath{\hat{b}^{\dagger}}\xspace}
	\newcommand{\boldr}{\ensuremath{\mathbf{r}}\xspace}
	\newcommand{\dr}{\ensuremath{\,d^3\mathbf{r}}\xspace}
	\newcommand{\tr}{\ensuremath{\,\mathrm{Tr}}\xspace}
	\newcommand{\dk}{\ensuremath{\,d^3\mathbf{k}}\xspace}
	\newcommand{\etal}{\emph{et al.\/}\xspace}
	\newcommand{\ie}{i.e.}
	\newcommand{\eq}[1]{Eq.~(\ref{#1})\xspace}
	\newcommand{\fig}[1]{Fig.~\ref{#1}\xspace}
	\newcommand{\abs}[1]{\left| #1 \right|}
	\newcommand{\proj}[2]{\left| #1 \rangle\langle #2\right| \xspace}
	\newcommand{\Qhat}{\ensuremath{\hat{Q}}\xspace}
	\newcommand{\Qhatd}{\ensuremath{\hat{Q}^\dag}\xspace}
	\newcommand{\phihatd}{\ensuremath{\hat{\phi}^{\dagger}}\xspace}
	\newcommand{\phihat}{\ensuremath{\hat{\phi}}\xspace}
	\newcommand{\boldk}{\ensuremath{\mathbf{k}}\xspace}
	\newcommand{\boldp}{\ensuremath{\mathbf{p}}\xspace}
	\newcommand{\boldsigma}{\ensuremath{\boldsymbol\sigma}\xspace}
	\newcommand{\boldalpha}{\ensuremath{\boldsymbol\alpha}\xspace}
	\newcommand{\grad}{\ensuremath{\boldsymbol\nabla}\xspace}
	\newcommand{\parti}[2]{\frac{ \partial #1}{\partial #2} \xspace}
	\newcommand{\vs}[1]{\ensuremath{\boldsymbol{#1}}\xspace}
	\renewcommand{\v}[1]{\ensuremath{\mathbf{#1}}\xspace}
	\newcommand{\Psihat}{\ensuremath{\hat{\Psi}}\xspace}
	\newcommand{\Psihatd}{\ensuremath{\hat{\Psi}^{\dagger}}\xspace}
	\newcommand{\Vhatd}{\ensuremath{\hat{V}^{\dagger}}\xspace}
	\newcommand{\Xhat}{\ensuremath{\hat{X}}\xspace}
	\newcommand{\Xhatd}{\ensuremath{\hat{X}^{\dag}}\xspace}
	\newcommand{\Yhat}{\ensuremath{\hat{Y}}\xspace}
	\newcommand{\Jhat}{\ensuremath{\hat{J}}\xspace}
	\newcommand{\Yhatd}{\ensuremath{\hat{Y}^{\dag}}\xspace}
	\newcommand{\jhat}{\ensuremath{\hat{J}}\xspace}
	\newcommand{\lhat}{\ensuremath{\hat{L}}\xspace}
	\newcommand{\Nhat}{\ensuremath{\hat{N}}\xspace}
	\newcommand{\rhohat}{\ensuremath{\hat{\rho}}\xspace}
	\newcommand{\ddt}{\ensuremath{\frac{d}{dt}}\xspace}
	\newcommand{\nset}{\ensuremath{n_1, n_2,\dots, n_k}\xspace}
	\newcommand{\Var}{\ensuremath{\mathrm{Var}}\xspace}
	\newcommand{\notes}[1]{{\color{blue}#1}}
	\newcommand{\cmc}[1]{{\color{red}#1}}
	\newcommand{\sah}[1]{{\color{magenta}#1}}
	\newcommand{\mxk}[1]{{\color{blue}#1}}
	\newcommand{\sss}[1]{{\color{blue}#1}}
	\newcommand{\mk}[1]{{\color{green}#1}}
	
	\newcommand{\hjz}{\hat{J}_z}
	\newcommand{\hsz}{\hat{S}_z}
	\newcommand{\kPT}{|\Psi(T)\rangle}
	\newcommand{\bPT}{\langle\Psi(T)|}
	\newcommand{\kPn}{|\Psi_0\rangle}
	\newcommand{\bPn}{\langle\Psi_0|}
	\newcommand{\kdgP}{|\partial_g\Psi(T)\rangle}
	\newcommand{\bdgP}{\langle\partial_g\Psi(T)|}
	\newcommand{\dgPdgP}{\langle\partial_g\Psi(T)|\partial_g\Psi(T)\rangle}
	\newcommand{\bPdgP}{\langle\Psi(T)|\partial_g\Psi(T)\rangle}
	\newcommand{\hp}{\hat{p}}
	\newcommand{\hG}{\hat{G}_0}
	\newcommand{\go}{\hG(T_1)}
	\newcommand{\gt}{\hG(T_2)}
	\newcommand{\hge}{\hat{G}_e}
	\newcommand{\hup}{\hat{U}_\pi}
	\newcommand{\ego}{e^{-ig\hG(T_1)}}
	\newcommand{\egt}{e^{-ig\hG(T_2)}}
	\newcommand{\eg}{e^{-ig\hG(T)}}
	\newcommand{\egd}{e^{ig\hG(T)}}
	\newcommand{\ege}{e^{-ig\hge}}
	\newcommand{\eko}{e^{-i\frac{T_1}{\hbar}\frac{\hat{\textbf{p}}^2}{2m}}}
	\newcommand{\ekt}{e^{-i\frac{T_2}{\hbar}\frac{\hat{\textbf{p}}^2}{2m}}}
	\newcommand{\ekto}{e^{-i\frac{T}{\hbar}\frac{\hat{\textbf{p}}^2}{2m}}}
	\newcommand{\hb}{\hbar}
	\newcommand{\hub}{\hat{U}_{\frac{\pi}{2}}}
	\newcommand{\ekzd}{e^{-ik_0\hat{z}}}
	\newcommand{\ekz}{e^{ik_0\hat{z}}}
	\newcommand{\ka}{|a\rangle}
	\newcommand{\ba}{\langle a|}
	\newcommand{\kb}{|b\rangle}
	\newcommand{\bb}{\langle b|}
	\newcommand{\hui}{\hat{U}_\text{int}}
	\newcommand{\hue}{\hat{U}_\text{ext}}
	\newcommand{\hued}{\hat{U}_\text{ext}^\dagger}
	\newcommand{\kpn}{|\psi_0\rangle}
	\newcommand{\bpn}{\langle\psi_0|}
	\newcommand{\hz}{\hat{z}}
	\newcommand{\hpz}{\hat{p}_z}
	\newcommand{\kPaT}{|\Psi_a(T)\rangle}
	\newcommand{\bPaT}{\langle\Psi_a(T)|}

\graphicspath{{figures_draft2/}} 
	
\title{Optimal Matterwave Gravimetry}
\author{Michail Kritsotakis}
\affiliation{Department of Physics and Astronomy, University of Sussex, Brighton BN1 9QH, United Kingdom}
\email{M.Kritsotakis@sussex.ac.uk}
\author{Stuart S.~Szigeti}
\affiliation{Department of Quantum Science, Research School of Physics and Engineering, The Australian National University, Canberra ACT 2601, Australia}
\affiliation{Department of Physics, Centre for Quantum Science, and Dodd-Walls Centre for Photonic and Quantum Technologies, University of Otago, Dunedin 9010, New Zealand}
\author{Jacob A.~Dunningham}
\affiliation{Department of Physics and Astronomy, University of Sussex, Brighton BN1 9QH, United Kingdom}
\author{Simon A.~Haine}
\affiliation{Department of Physics and Astronomy, University of Sussex, Brighton BN1 9QH, United Kingdom}
\affiliation{Department of Quantum Science, Research School of Physics and Engineering, The Australian National University, Canberra ACT 2601, Australia}

\begin{abstract}
We calculate quantum and classical Fisher informations for gravity sensors based on matterwave interference, and find that current Mach-Zehnder interferometry is not optimally extracting the full metrological potential of these sensors. We show that by making measurements that resolve either the momentum or the position we can considerably improve the sensitivity. We also provide a simple modification that is capable of more than doubling the sensitivity.
\end{abstract}

\maketitle

Atom interferometry is a leading inertial-sensing technology, having demonstrated state-of-the-art gravimetry \cite{Peters:1999, Peters:2001, Hu:2013, Hauth:2013, Altin:2013, Freier:2016, Hardman:2016b} and gradiometry \cite{Snadden:1998, Stern:2009, Sorrentino:2014, Biedermann:2015, DAmico:2016, Asenbaum:2017, Bidel:2018} measurements. Nevertheless, orders of magnitude improvement in sensitivity is required for applications in navigation \cite{Richeson:2008} and mineral exploration \cite{Evstifeev:2017}, as well as improved tests of the equivalence principle \cite{Fray:2004,Dimopoulos:2007,Schlippert:2014} and quantum gravity \cite{Amelino-Camelia:2009, Gao:2016}. For the commonly-used Mach-Zehnder [i.e.~Kasevich-Chu (KC)] configuration \cite{Borde:1989, Kasevich:1991}, semiclassical calculations \cite{Kasevich:1992, Storey:1994, Schleich:2013, Schleich:2013b} reveal that the matterwave accrues relative phase $\phi = \textbf{g} \cdot \textbf{k}_L T_\pi^2$, where $\textbf{g}$ is the gravitational acceleration, $\hbar \textbf{k}_L$ is the momentum separation of the two arms, and $2 T_\pi$ is the total interrogation time. Assuming $N$ uncorrelated particles, a population-difference measurement at the interferometer output yields sensitivity
\begin{equation}
	\Delta g = \frac{1}{\sqrt{N} k_0 T_\pi^2}, \label{delta_g_KC} 
\end{equation}
where $k_0$ is the component of $\textbf{k}_L$ aligned with $\textbf{g}$. Equation~(\ref{delta_g_KC}) implies only four routes to improved sensitivity: (1) increase interrogation time, (2) increase the momentum separation of the arms (e.g. via large momentum transfer beam splitters \cite{Muller:2008b, Clade:2009, Chiow:2011, McDonald:2013b, Mazzoni:2015}), (3) increase the atom flux, and/or (4) surpass the shot-noise limit with quantum correlations \cite{Gross:2010, Lucke:2011, Linnemann:2016, Hosten:2016,Colangelo:2017}. Although all routes are worth pursuing, each has unique limitations. For instance, size, weight, and power constraints limit both $T_\pi$ and the maximum momentum transferrable via laser pulses. Additionally, evaporative-cooling losses and momentum width requirements constrain atom fluxes \cite{Debs:2011, Louchet-Chauvet:2011, Szigeti:2012, Hardman:2014, Robins:2013}. Increases with number-conserving feedback cooling are possible, but untested~\cite{Szigeti:2009, Szigeti:2010, Hush:2013}. Finally, quantum-correlated states must be compatible with the requirements of high-precision metrology \cite{Altin:2011, Haine:2011, Altin:2013, Haine:2013, Haine:2014, Szigeti:2014b, Tonekaboni:2015, Haine:2015, Haine:2015b, Nolan:2016, Haine:2016, Szigeti:2017, Haine:2018} (e.g. high atom flux, low phase diffusion), and will only be advantageous if classical noise sources (e.g. \cite{Lan:2012, Schkolnik:2015}) are sufficiently controlled to yield shot-noise-limited operation \emph{prior} to quantum enhancement. 

This assessment assumes that Eq.~(\ref{delta_g_KC}) is the \emph{optimal} sensitivity. In this article, we prove this conventional wisdom false by showing that matterwave interferometers \emph{can} attain better sensitivities than Eq.~(\ref{delta_g_KC}). Ultimately, the gravitational field affects the quantum state beyond the creation of a simple phase shift. We show this additional metrological potential via the quantum Fisher information (QFI), which determines the \emph{best} possible sensitivity. We further determine the set of measurements required to attain this optimal sensitivity via the classical Fisher information (CFI). Our analysis reveals additional routes to improved sensitivity, such as variations in the measurement procedure and input source, and these should be considered when designing future matterwave gravimeters. We also present a modified interferometer that more than doubles the sensitivity for the same interrogation time and momentum separation.

\begin{figure}[h]
	\begin{center}
		\includegraphics[width=\columnwidth]{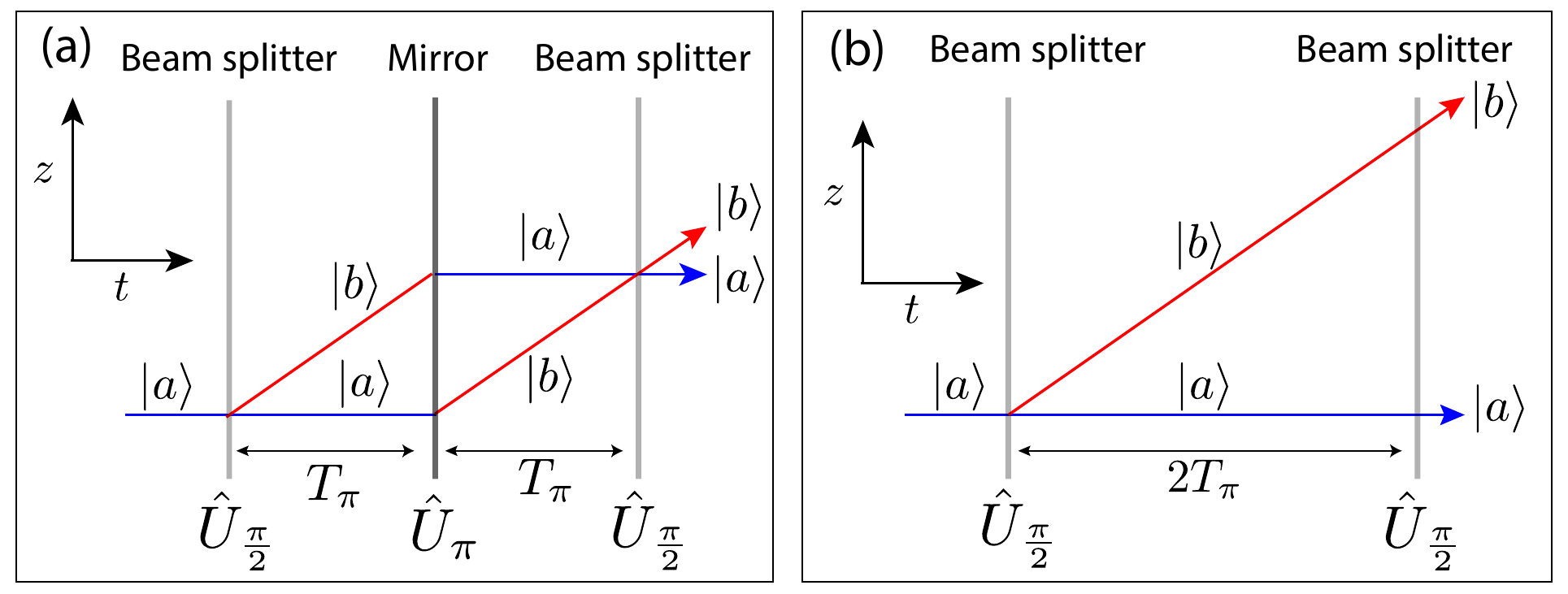}
		\caption{ Spacetime diagrams for (a) KC interferometry and (b) Ramsey interferometry (no mirror pulse), which are both sensitive to gravitational fields and accelerations.}
		\label{fig:scheme}
	\end{center}
\end{figure}

The focus of this article is KC interferometry based on state-changing Raman transitions, although our results also hold for Bragg transitions \cite{Altin:2013} and Bloch oscillations \cite{Clade:2009} in the appropriate regime. A KC interferometer is schematically depicted in Fig.~\ref{fig:scheme}(a). At time $t=0$ atoms with two internal states $|a\rangle$ and $|b\rangle$, initially in $|a\rangle$, are excited to an equal superposition of $|a \rangle$ and $|b \rangle$ via a coherent $\pi/2$ pulse. Atoms transferred to $|b\rangle$ also receive a momentum kick $\hbar k_0$. 
At $t= T_\pi$, a $\pi$ pulse acts as a mirror, before the two matterwaves are interfered at $t=T = 2T_\pi$ by a second $\pi/2$ pulse. 

\section{QFI for a particle in a gravitational field}
The quantum Cram\'er-Rao bound (QCRB) gives a lower bound on the sensitivity \cite{Braunstein:1994}. For $N$ uncorrelated particles this is $\Delta g^2 \geq 1 / (N F_Q)$, where $F_Q$ is the single-particle QFI \cite{Demkowicz-Dobrzanski:2014, Toth:2014, Haine:2016b}, which for a pure single-particle state $|\Psi \rangle$ is 
\begin{equation}
	F_Q =4\left(\langle \partial_g \Psi | \partial_g \Psi\rangle - \abs{\langle \Psi| \partial_g \Psi\rangle}^2\right). \label{general_QFI_equation}
\end{equation}

For the KC interferometer, semiclassical arguments give $|\Psi\rangle = \tfrac{1}{\sqrt{2}}(|a\rangle + e^{i g k_0 T^2}|b \rangle)$ before the final beam splitter and a QFI $F_Q^\text{sc} = k_0^2 T_\pi^4$ \cite{Qvarfort:2017}, consistent with \eq{delta_g_KC}. However, this derivation treats the particle's motion semiclassically, neglecting the noncommutability of position and momentum. We account for this here. For the moment we consider only the centre of mass degrees of freedom. In the presence of a uniform gravitational field $g$ acting along the $z$-axis, a particle of mass $m$ in state $|\psi_0 \rangle$ evolves to $|\psi(T)\rangle = \hat{U}_g |\psi_0\rangle$ after time $T$, where $\hat{U}_g = \exp[-\frac{iT}{\hbar}(\frac{\hat{\boldp}^2}{2m} + mg\hat{z} ) ]$. As shown in Appendix~\ref{appendix_QFI_particle}, we can rewrite
\begin{equation}
	\hat{U}_g = \ekto\eg e^{i\frac{mg^2T^3}{12\hbar}},
\end{equation}
where
\begin{equation}
	\hat{G}_0(T) = \tfrac{T}{\hbar}\left(\tfrac{T}{2}\hat{p}_z + m\hat{z}\right).
\end{equation}
The QFI is
\begin{subequations}
\begin{align}
&F_Q(T) = 4\mathrm{Var}(G_0(T)) \label{FQ_grav} \\ 
&= \tfrac{T^4}{\hbar^2}\Var(p_z) + \tfrac{4m^2 T^2}{\hbar^2} \Var(z) + \tfrac{4m T^3}{\hbar^2} \mathrm{Cov}(p_z, z), \label{FQ_grav_terms}
\end{align}
\end{subequations}
where the variances and covariance are evaluated with respect to $|\psi_0\rangle$.
To compare \eq{FQ_grav} and $F_Q^\text{sc}$, consider a state $|\psi_0\rangle$ with two well-defined peaks in momentum space separated by $\hbar k_0$, giving $\mathrm{Var}(p_z) \approx (\hbar k_0)^2$. For sufficiently large $k_0$ and $T$ such that $(\hbar k_{0} T/2)^2 \gg m^2 \mathrm{Var}(z)$, $mT\mathrm{Cov}(p_z,z)$, the first term of Eq.~(\ref{FQ_grav_terms}) dominates, and $F_Q(2T_\pi)\approx k_0^2 T_\pi^4 = F_Q^\text{sc}$. However, the additional terms in Eq.~(\ref{FQ_grav_terms}) potentially allow sensitivities better than \eq{delta_g_KC}. 


\section{QFI for KC interferometry}
Equation~(\ref{FQ_grav}) is \emph{not} the QFI for a KC interferometer, as we must account for the internal state degrees of freedom as well as the action of the mirror pulse. The evolution is given by 
\begin{equation}
	\hat{U}_\text{KC} = \hat{U}_{\frac{\pi}{2}}^{\phi_3} \hat{U}_g(T_2)\hat{U}_{\pi}^{\phi_2}\hat{U}_g(T_1)\hat{U}_{\frac{\pi}{2}}^{\phi_1},
\end{equation}
where
\begin{equation}
	\hat{U}_{\theta}^\phi = \hat{1}\cos\left(\tfrac{\theta}{2}\right) - i(|b\rangle\langle a| e^{i(k_0 \hat{z}-\phi)} + \mathrm{h.c.}) \sin\left(\tfrac{\theta}{2}\right) \label{BS_unitary}
\end{equation}
governs the beam splitter and mirror dynamics. As shown in Appendix~\ref{appendix_BS_derivation}, Eq.~(\ref{BS_unitary}) is an excellent approximation to the beam splitting and mirror dynamics when the pulse duration is much shorter than the timescale for atomic motional dynamics. Here $T_{1(2)}$ are evolution times before(after) the $\pi$ pulse and $\phi$ is the pulse phase, controlled via the relative phase of the two Raman lasers. The first $\pi/2$ pulse maps the initial state $|\Psi_0\rangle = |a\rangle |\psi_0\rangle $ to $|\Psi^\prime_0\rangle = \hat{U}_{\frac{\pi}{2}}^{\phi_1}|\Psi_0\rangle = \tfrac{1}{\sqrt{2}}\left(|a\rangle - i e^{i (k_0 \hat{z}-\phi_1)} |b\rangle\right) |\psi_0\rangle $, where $|\psi_0\rangle$ contains the initial state's motional degrees of freedom. As detailed in Appendix~\ref{appendix_QFI_KC},
\begin{align}
	|\Psi(T)\rangle 	&= \hat{U}_\text{KC} |\Psi_0 \rangle = \hat{U}_{0}e^{-ig(\hat{G}_0(T) + \hat{G}_e)} |\Psi_0^\prime\rangle, \label{psiT_KC_simp}
\end{align} 
where 
\begin{subequations}
	\begin{align}
		\hat{G}_e  &= \hat{S}_z k_0 T_2^2, \\
		\hat{S}_z 	&= \frac{1}{2} \left(|a\rangle\langle a| - |b\rangle\langle b|\right), \\
		\hat{U}_0 	&= \hat{U}_{\frac{\pi}{2}}^{\phi_3} e^{ -i \frac{T_2}{\hbar} \frac{\hat{\boldp}^2}{2m}} \hat{U}_{\pi}^{\phi_2} e^{ -i\frac{T_1}{\hbar} \frac{\hat{\boldp}^2}{2m}},
	\end{align}
\end{subequations}
and $T= T_1 +T_2$, giving QFI
\begin{equation}
F_Q^\text{KC}(T) = 4 \mathrm{Var}(G_0(T))  + \tfrac{1}{4}k_0^2\left(T^2 - 2T_2^2\right)^2, \label{FQKC}
\end{equation}
where $\mathrm{Var}(G_0(T))$ is taken with respect to $|\psi_0\rangle$. For $T_1=T_2 = T_\pi$,
\begin{equation}
	F_Q^\text{KC}(T) = 4 \mathrm{Var}(G_0(T))  + k_0^2T_\pi^4. \label{FQ_full}
\end{equation}	
Since $\mathrm{Var}(G_0(T)) \geq 0$, this implies $F_Q^\text{KC} \geq F_Q^\text{sc}$, thereby permitting sensitivities better than \eq{delta_g_KC}. 

\section{Classical Fisher information}
Although the QFI gives the best possible sensitivity, it is silent on how to \emph{achieve} this sensitivity. The attainable sensitivity for a particular measurement choice is given by the CFI, which quantifies the information contained in the probability distribution constructed from measurements of a particular observable, and necessarily depends upon this choice of observable. We calculate the CFI via 
\begin{equation}
	F_C(\hat{\Lambda}) = \int d\lambda \frac{[\partial_g P(\lambda)]^2}{P(\lambda)},
\end{equation}
where $P(\lambda)$ is the probability of obtaining result $\lambda$ when the observable $\hat{\Lambda}$ is measured \cite{Demkowicz-Dobrzanski:2014, Toth:2014}. The CFI is bounded by the QCRB $F_C \leq F_Q$, so a measurement that saturates this bound is the optimal measurement. 

\subsection{CFI for population-difference measurement}
For the standard population-difference measurement at the KC interferometer output, $\hat{\Lambda} = \hat{S}_z$ and $F_C(\hat{S}_z) = \sum_{s=a,b} (\partial_g P_s)^2 / P_s $, where $P_s =  \int dz | \langle s| \langle z|\Psi(T)\rangle |^2 $. As detailed in Appendix~\ref{appendix_CFI_KC}, an analytic solution exists in this case. Specifically,
\begin{subequations}
\label{CFI_probs_equations}
\begin{align}
	P_a 	&= \tfrac{1}{2}(1+|\mathcal{C}|\sin\alpha ), \\
	P_b	&= \tfrac{1}{2}\left(1-|\mathcal{C}|\sin\alpha\right),
\end{align}
\end{subequations}
yielding
\begin{equation}
	F_C(\hat{S_z})=\frac{|\mathcal{C}|^2\cos^2\alpha}{1-|\mathcal{C}|^2\sin^2\alpha}k_0^2\left(\tfrac{T^2}{2}-T_1^2\right)^2,  \label{FC_anal}
\end{equation}
where 
\begin{subequations}
\begin{align}
	\mathcal{C}	&=\langle\psi_0|e^{i\frac{k_0}{m}(T_2-T_1)\hat{p}_z}|\psi_0\rangle \equiv |\mathcal{C}|e^{i\vartheta}, \\
	\alpha		&=\phi_f-\phi_g+\vartheta,
\end{align} 
\end{subequations}
with $\phi_f = \frac{\hbar k_0^2}{2m}(T_2-T_1)$ and $\phi_g = k_0 g (\frac{T^2}{2}-T_1^2 )$. The contrast $|\mathcal{C}|$ is determined by the spatial overlap of the two output wavepackets, since $\frac{\hbar k_0}{m}(T_2-T_1)$ is the spatial separation. This depends strongly on the time difference $T_2-T_1$. For an initial Gaussian state $\langle z|\psi_0\rangle = \exp(-z^2/2\sigma^2) / (\pi \sigma^2)^{1/4}$, $|\mathcal{C}|=\exp[-\frac{\hbar^2k_0^2}{4m^2\sigma^2}(T_2-T_1)^2]$.

Figure~\ref{FI_ho}(a) shows the time dependence of the QFI and $F_C(\hat{S}_z)$ for this initial Gaussian state. Here $t = T_1 + T_2$, we fix $T_\pi$ so the mirror pulse always occurs at $t = T_\pi$, and the second beam splitter occurs instantaneously before measurement. Explicitly, if $t \leq T_\pi$, then $T_1 = t$, $T_2 = 0$, and the mirror pulse has no meaningful effect; if $t > T_\pi$ then $T_1 = T_\pi$ and $T_2 = t - T_\pi$. When $T_1$ and $T_2$ are significantly different, the spatial overlap of the two modes at the interferometer output is poor, so both the contrast and CFI are close to zero. However, $|\mathcal{C}| = 1$ when $T_1 = T_2$ and $F_C(\hat{S}_z) = F_Q^\text{sc}=k_0^2T_\pi^4$, giving the same sensitivity as \eq{delta_g_KC}. This is still less than $F_Q^\text{KC}$, indicating that a different measurement could yield improved sensitivities.

\subsection{CFI for momentum-distribution measurement}
Now consider a measurement that distinguishes internal states \emph{and} fully resolves the $z$-component of the final momentum distribution, such as reported in Ref.~\cite{Matthews:1999}. This measurement yields CFI
\begin{equation}
	F_C( \hat{S}_z,\hat{p}_z ) = \sum_{s=a,b}\int dp_z \frac{[\partial_g P_s(p_z)]^2}{P_s(p_z)},
\end{equation} 
where $P_s(p_z) = \abs{\langle s|\langle p_z|\Psi(T)\rangle}^2$. Although no analytic formula exists for $F_C(\hat{S}_z,\hat{p}_z)$, the probabilities can be determined by numerically solving the Schr\"odinger equation, and  the CFI computed from finite differences of these probabilities \cite{Haine:2016b}. This requires an explicit choice of $g$; although we consider the sensitivity near $g = 0$ for all numerical calculations, a large offset in $g$ is easily accounted for by adjusting the beam splitter phases, as in typical atomic gravimeters \cite{Hardman:2014}.

\begin{figure}[t!]
	\begin{center}
		\includegraphics[width=\columnwidth]{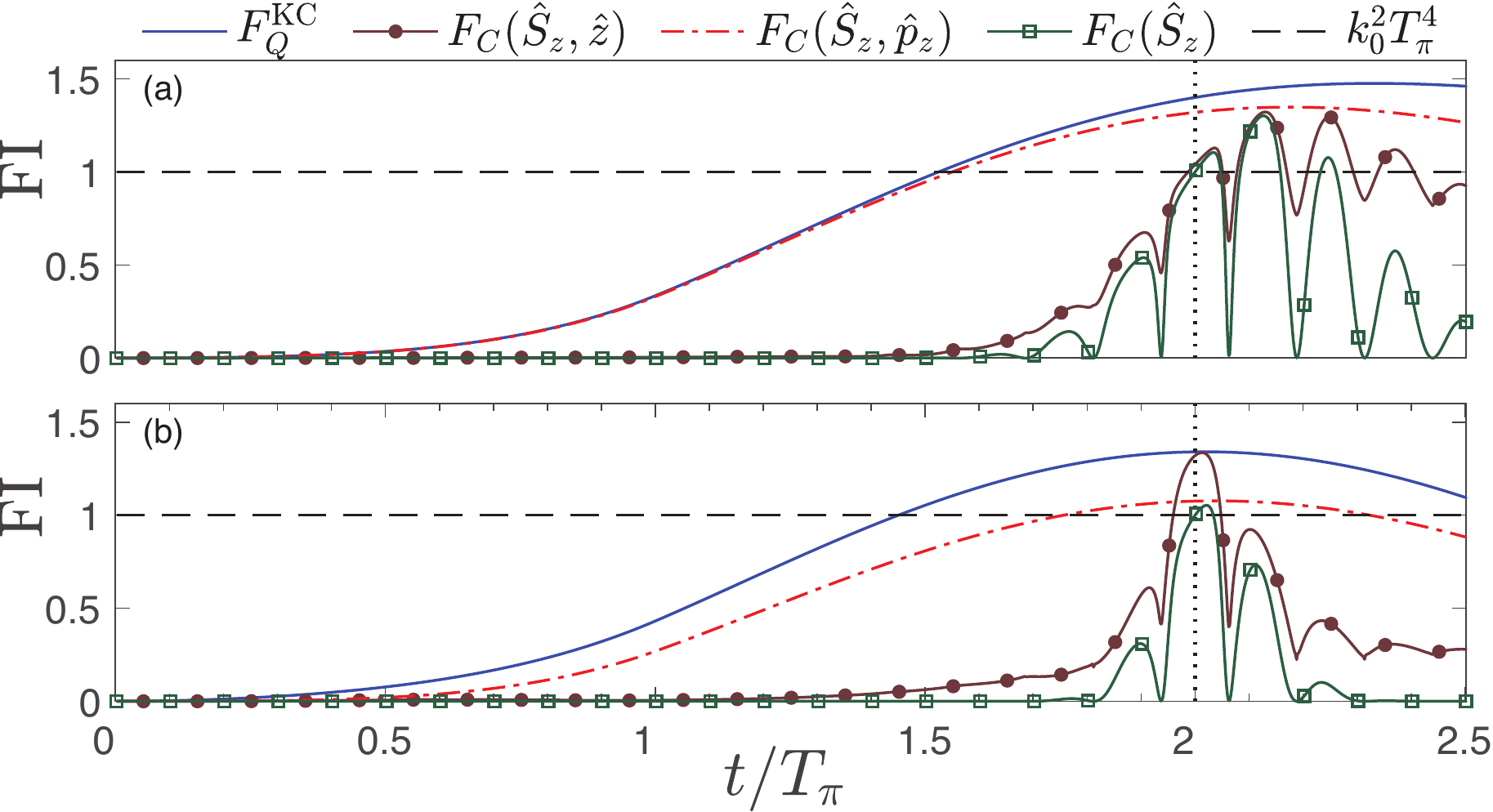}
\caption{Fisher information (FI) for $|\Psi(t)\rangle = \hat{U}_\text{KC}(t)|\Psi_0\rangle$, where $T_1 = t$ and $T_2 = 0$ for $t \leq T_\pi$, otherwise $T_1 = T_\pi$ and $T_2 = t- T_\pi$, with initial Gaussian motional state (a) $\langle z|\psi_0\rangle = \exp(-z^2/2\sigma^2) / (\pi \sigma^2)^{1/4}$ and (b) $\langle z|\psi_0\rangle = e^{-(\frac{1}{4}+i)z^2/2\sigma^2} / [\pi (2 \sigma)^2]^{1/4}$. FI has units $k_0^2 T_\pi^4$, so when FI~$> 1$ a given measurement scheme achieves a sensitivity better than that predicted by the semiclassical limit Eq.~(\ref{delta_g_KC}). The QFI $F_Q^\text{KC}$ gives the maximum possible FI. Here $\sigma = 10 L$ and $T_\pi = 100 t_0$, whilst the length ($L = k_0^{-1}$) and time ($t_0 = m/\hbar k_0^2$) units depend on $k_0$.}
\label{FI_ho}
	\end{center}
\end{figure}

Figure~\ref{FI_ho}(a) shows that $F_C(\hat{S}_z, \hat{p}_z)$ is significantly larger than $F_C(\hat{S}_z)$ and very close to $F_Q^\text{KC}$. Additionally, $F_C(\hat{S}_z, \hat{p}_z) \approx F_Q^\text{KC}$ even when $T_1$ and $T_2$ are vastly different. This is because $P_s(p_z)$ displays interference fringes that are not present in $P_s = \int dp_z P_s(p_z)$ when spatial overlap is poor.

The origin of the increased information in $F_C(\hat{S}_z, \hat{p}_z)$ compared with $F_C(\hat{S}_z)$ is easily understood. Additional to the CFI associated with population exchange (generated by $\hat{G}_e$), there is information due to a shift in the momentum distribution. Concretely, consider initial momentum distribution $P_0(p_z)$. Under gravity, $\hat{p}_z(t) = \hat{p}_z(0) + mgt$, so $P(p_z,t) = P_0(p_z - mgt)$, giving 
\begin{align}
	F_C(p_z) 	&= \int dp_z \frac{[\partial_g P(p_z,t)]^2}{P(p_z,t)} \notag \\
			& = [\partial_g p_z(t)]^2 \int dp_z \frac{[ \partial_{p_z} P_0(p_z)]^2}{ P_0(p_z)} \notag \\
			& \equiv (m t)^2 F^{p_z}_{C}, 
\end{align}
where $F_C^{p_z}$ is the CFI associated with resolvable small shifts in the momentum distribution. For the initial Gaussian considered in Fig.~\ref{FI_ho}(a), adding this additional CFI to $F_C(\hat{S}_z)$ gives $F_C(\hat{S}_z, \hat{p}_z)|_{2T_\pi} = F_Q^\text{sc} +  8(m T_\pi \sigma/\hbar)^2$, in perfect agreement with our numerics. Note that this additional information is \emph{not} the result of a phase shift so, unlike a standard KC interferometer, it is \emph{not} affected by additional phase noise. 

Our simulations also find near-perfect correlations between internal and momentum states, so a measurement that only resolves momentum (and not $\hat{S}_z$) also has CFI approximating $F_C(\hat{S}_z, \hat{p}_z)|_{2T_\pi}$, since an atom's internal state is inferred from its final momentum.  Our analysis therefore holds for interferometers that do not change internal states, such as Bragg-scattering-based interferometers, provided $\hbar k_0 \gg \delta p$, where $\delta p$ is the wave packet's initial momentum width \cite{Muller:2008b, Altin:2013}. In our simulations $\hbar k_0\approx  14\delta p$. 

\subsection{CFI for position-distribution measurement}
Although the momentum distribution cannot always be resolved, a measurement of the \emph{position} distribution might be possible. Here the CFI is 
\begin{equation}
	F_C(\hat{S}_z,\hat{z}) = \sum_{s=a,b}\int dz \frac{[\partial_g P_s(z)]^2}{P_s(z)},
\end{equation} 
where $P_s(z) = \abs{\langle s| \langle z| \Psi(t)\rangle}^2$. Figure~\ref{FI_ho}(a) shows this is slightly better than the population-difference measurement, although significantly worse than the momentum measurement. Arguing as before, since the position distribution shifts due to $\hat{z}(t) = \hat{z}(0) + \hat{p}_z(0)t / m + \frac{1}{2}g t^2$, the additional CFI is $(t^2/2)^2 F_C^z$, where $F_C^z = \int dz [\partial_z P(z)]^2/P(z)$ is the CFI associated with resolvable shifts in the position distribution. Since 
\begin{align}
	\mathrm{Var}(z(t)) 	&= \mathrm{Var}(z(0)) + \frac{t^2}{m^2} \mathrm{Var}(p_z(0)) \notag \\
					& + \frac{t}{2m}\mathrm{Cov}(p_z(0), z(0)),
\end{align} 
and $F_C^z = 1/\Var(z)$ for Gaussian states, we obtain $F_C(\hat{S}_z, \hat{z})|_{2T_\pi} = F_Q^\text{sc} + 8(\sigma m T_\pi^2)^2/[(\sigma^2 m)^2 + (2\hbar T_\pi)^2]$ for the initial Gaussian considered in Fig.~\ref{FI_ho}(a), in agreement with numerics.

We can increase $F_C(\hat{S}_z,\hat{z})$ with an initial state that decreases $\Var(z(2T_\pi))$ at the interferometer output. This is not achieved by reducing $\Var(z(0))$, but rather via an initial state with nontrivial correlations between position and momentum such that $\mathrm{Cov}(\hat{p}_z, \hat{z})$ counteracts the wave packet's ballistic expansion. Figure~\ref{FI_ho}(b) shows the QFI and CFI for initial state $\langle z|\psi_0\rangle = e^{-(\frac{1}{4}+i)z^2/2\sigma^2} / [\pi (2 \sigma)^2]^{1/4}$. The imaginary term provides the position-momentum correlations and doubling the spatial width increases the ability of the wavepacket to be \emph{focused}. This initial state could be engineered by applying a harmonic potential for a short duration (compared to motional dynamics), creating phase gradient $\psi(z) \rightarrow \psi(z)e^{-i z^2/\sigma_t^2}$, for constant $\sigma_t$ which depends on trap frequency and duration \cite{Ammann:1997}. Then $F_C(\hat{S}_z,\hat{z})$ saturates the QCRB at $T_1 = T_2$, at the cost of reduced $F_C(\hat{S}_z,\hat{p})$.

\section{Optimum measurements}
Since measurements in different bases yield different sensitivities, is there an accessible measurement basis that saturates the QCRB? Our above analysis suggests yes and, depending on the initial state, this optimum basis lies somewhere \emph{between} position and momentum. We confirm this intuition by revisiting a particle in a gravitational field. We rewrite 
\begin{equation}
	|\psi(t)\rangle = \hat{U}_g|\psi_0\rangle = \exp(-i g \hat{G}_0^\prime(t) )|\psi_0(t)\rangle,
\end{equation}
where 
\begin{align}
	\hat{G}_0^\prime(t) 	&= \hat{U}_p\hat{G}_0(t)\hat{U}_p^\dag = \frac{t}{\hbar} (m\hat{z} - \tfrac{1}{2}\hat{p}_z t), 
\end{align}
$\hat{U}_p = \exp[-i t \boldp^2/(2m\hbar)]$, and $|\psi_0(t)\rangle = \hat{U}_p|\psi_0\rangle$ describes free-particle evolution. We can interpret $\hat{G}_0'(t)$ as the generator of displacements in $\hat{Q} = c_1 \hat{z} + c_2 \hat{p}_z$, where the coefficients $c_{i}$ are real and chosen such that $[\hat{G}_0^\prime(t), \hat{Q}] = i$. Hence, the probability distribution $|\langle q |\psi(t)\rangle|^2 = |\langle q - g| \psi_0(t)\rangle|^2$,
where $\hat{Q}|q\rangle = q | q \rangle$. If $|\langle q |\psi_0(t)\rangle |^2$ is Gaussian, then measurements of $\hat{Q}$ saturate the QCRB, since $[\hat{G}_0^\prime(t), \hat{Q}] = i$ implies 
\begin{equation}
	F_C(\hat{Q}) = \frac{1}{\Var (Q)} = 4\Var(G_0^\prime(t)) = F_Q.
\end{equation}

To measure $\hat{Q}$, we mix $\hat{z}$ and $\hat{p}_z$ by applying the potential $V(z) = \tfrac{1}{2}m\omega^2 z^2$, since $\hat{z}(t) = \hat{z}(0) \cos \omega t + [\hat{p}(0)/m\omega] \sin \omega t$. Subsequently measuring position yields a combination of position and momentum information. This scheme could be implemented using the following procedure:
\begin{enumerate}
	\item At $t=2 T_\pi$, apply the unitary $\hat{U}_s = |a\rangle\langle a| + |b\rangle\langle b| e^{-ik_0 \hat{z}}$, which removes any momentum mismatch between the two modes. A state-selective Bragg transition achieves this.
	\item Then apply the potential $V(z) = \frac{1}{2}m\omega^2 (z-z_0)^2$, where $z_0 = \hbar k_0 T_\pi/m$ is the matterwave's centre-of-mass displacement at the interferometer output.
	\item Finally, at some later time, we apply a beam splitter $\hat{U}_\text{BS} = \frac{1}{\sqrt{2}}[ \hat{1} + (|a\rangle\langle b| - \mathrm{h.c.})]$ immediately before measurement.
\end{enumerate}
%
\begin{figure}[t!]
	\begin{center}
		\includegraphics[width=\columnwidth]{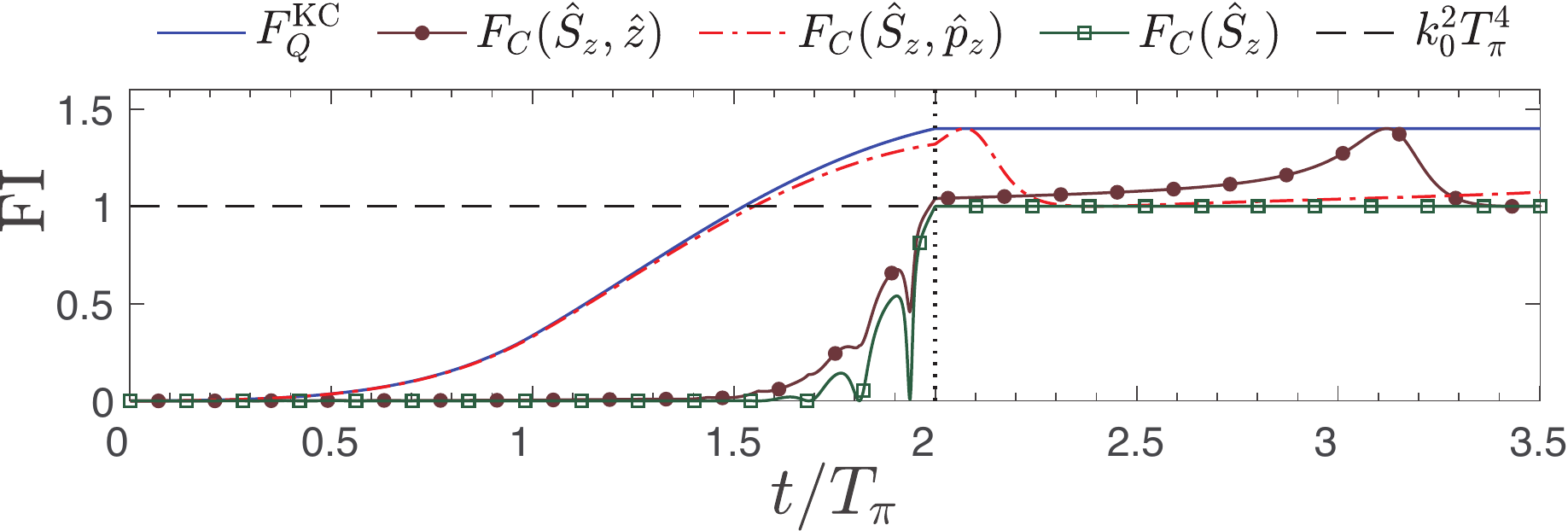}
\caption{Fisher information (FI) $|\Psi(t)\rangle = \hat{U}_\text{KC}(t)|\Psi_0\rangle$, where $T_1 = t$ and $T_2 = 0$ for $t \leq T_\pi$, otherwise $T_1 = T_\pi$ and $T_2 = t- T_\pi$, with a harmonic potential applied at $t=2T_\pi$ and initial Gaussian motional state $\langle z|\psi_0\rangle = \exp(-z^2/2\sigma^2) / (\pi \sigma^2)^{1/4}$. We artificially turned off gravity at $t = 2 T_\pi$ (which holds $F_Q^\text{KC}$ constant) to clearly show the effect of harmonic trapping. Specifically, the application of this harmonic potential can be used to saturate the QCRB with either a position-distribution or momentum-distribution measurement. Here $\sigma = 10 L$, $T_\pi = 100 t_0$, and $\omega = 3 \pi / (2T_\pi)$. FI has units $k_0^2 T_\pi^4$, and length ($L = k_0^{-1}$) and time ($t_0 = m/\hbar k_0^2$) units depend on $k_0$.}
\label{FI_ho_c}
	\end{center}
\end{figure}
Figure~\ref{FI_ho_c} shows $F_C(\hat{S}_z,\hat{z})$ and $F_C(\hat{S}_z,\hat{p}_z)$ for this scheme. Both CFIs oscillate between $F_Q^\text{sc}$ and the QFI, so a measurement in either the position or momentum basis saturates the QCRB if made at the appropriate time. This improved sensitivity \emph{does} increase the interferometer time. However, the period of CFI oscillations is negligible compared to $T_\pi$ for sufficiently large $\omega$.

\section{Improved interferometry}
In KC interferometry, the $\pi$ pulse ensures that the wavepackets spatially overlap at $t=2T_\pi$. However, Fig.~\ref{FI_ho} and Fig.~\ref{FI_ho_c} reveal that spatial overlap is \emph{not} required for a momentum measurement, making the mirror pulse unnecessary. More interestingly, removing the $\pi$ pulse significantly increases the spatial separation, and therefore the QFI, for the same interrogation time. More precisely, setting $T_1 = 2T_\pi$ and $T_2 = 0$ in \eq{FQKC} gives $F_Q(T) = 4\mathrm{Var}(G_0(T)) + 4 k_0^2T_\pi^4$, an increase of $3 F_Q^\text{sc}$ over symmetric KC interferometry. 

\begin{figure}[t!]
	\begin{center}
		\includegraphics[width=\columnwidth]{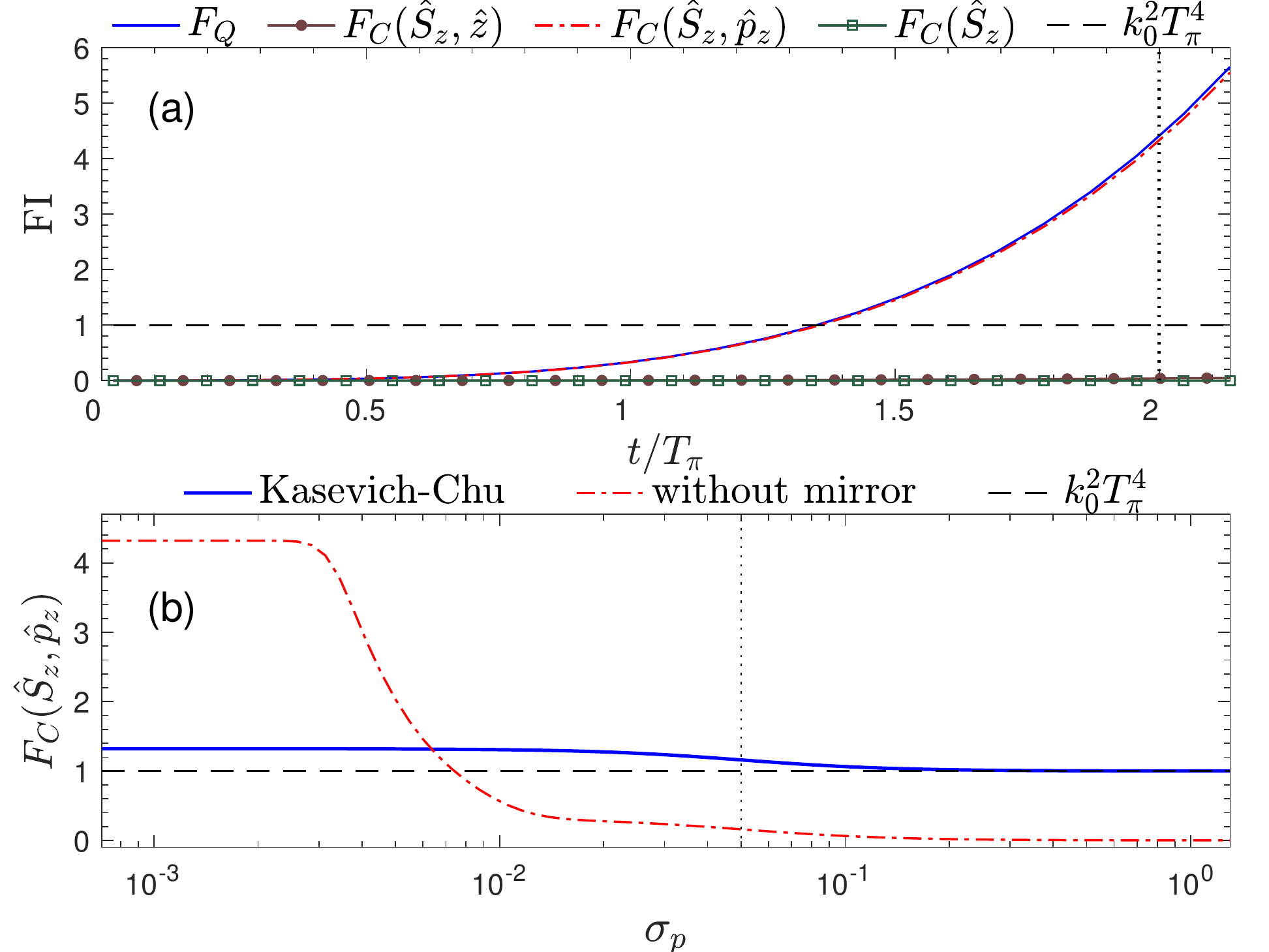}
		\caption{(a) FI of the mirrorless configuration for the same initial state and parameters as Fig.~\ref{FI_ho}(a). We normalize time by $T_\pi = 100 t_0$ only for comparison with Fig.~\ref{FI_ho}. Note that $F_C(\hat{S}_z)$ and $F_C(\hat{S}_z,\hat{z})$ are almost zero throughout the entire evolution, since there is no spatial overlap of the wavepackets and consequently no interference in $P_s$ or the position distribution. (b) $F_C(\hat{S}_z, \hat{p}_z)$ constructed from convolving probabilities with a Gaussian of width $\sigma_p$ (units $\hbar k_0$). The vertical line marks the initial state's momentum width: $\delta p = \hbar/\sqrt{2}\sigma \approx 0.07 \hbar k_0$. FI is in units of $k_0^2 T_\pi^4$.}
		\label{mirrorless_conv}
	\end{center}
\end{figure}

We numerically solved the Schr\"odinger equation for the mirrorless Mach-Zehnder (i.e.~Ramsey) configuration [Fig.~\ref{fig:scheme}(b)]. Figure~\ref{mirrorless_conv}(a) shows that a momentum measurement is always nearly optimal, and at $t=2T_\pi$, $F_C(\hat{S}_z, \hat{p}_z)/F_Q^\text{sc} \approx 4.4$. Unfortunately, this improved sensitivity has a price: A lack of spatial overlap means that information is encoded in high-frequency interference fringes in the momentum distribution, requiring high-resolution momentum measurements. Following Refs.~\cite{Pezze:2013, Gabbrielli:2015, Nolan:2017b, Mirkhalaf:2018, Haine:2018b}, we model imperfect resolution by convolving the momentum distribution at $t=2T_\pi$ with a Gaussian of width $\sigma_p$ before constructing $F_C(\hat{S}_z, \hat{p}_z)$ [Fig.~\ref{mirrorless_conv}(b)]. This imperfect resolution may be due to limitations on the detection system, or other sources of classical noise. The mirrorless configuration is considerably more sensitive to imperfect momentum resolution than KC interferometry, where $F_C(\hat{S}_z, \hat{p}_z)$ begins to degrade only when $\sigma_p$ is comparable to the initial wavepacket's momentum width. Furthermore, in the limit of a ``bad" momentum measurement ($\sigma_p \rightarrow \infty$), the CFI goes to zero, whereas the CFI for KC interferometry approaches $F_Q^\text{sc}$. Nevertheless, if high-resolution measurements are available (or actively developed), as reported in Ref.~\cite{Gotlibovych:2014} for instance, our result suggests that pursuing a mirrorless configuration could yield substantial sensitivity gains. 

\section{Discussion and Outlook}
An important experimental consideration is achieving high-resolution momentum measurements. Time-of-flight imaging is a standard technique, where ballistic expansion converts the momentum distribution into a position distribution \cite{Ketterle:1999, Hardman:2016}. However, the expansion time needed for sufficient momentum resolution might be significantly longer than the interrogation time, in which case longer interrogation times are a better route to improved sensitivities. Bragg spectroscopy is perhaps a more promising approach \cite{Stenger:1999, Richard:2003}. 
 
Reference~\cite{Hardman:2016b} reports state-of-the-art gravimetry with a Bose-Einstein condensate (BEC), well-described by a pure motional state, and parameters: $\sigma = 40 \mu$m, $T_\pi = 130$ms,  $k_0 = 1.6\times 10^7$ m$^{-1}$, and $\delta p_z = 0.18 \hbar k_0$. We estimate that $4\Var(G_0(T))$ is $\sim 7\%$ of $F_Q^\text{sc}$, so there is little gain in making optimal measurements [Eq.~(\ref{FQ_full})]. However, $4\Var(G_0(T)) \sim F_Q^\text{sc}$ if $\sigma$ or $\delta p_z$ were increased by an order of magnitude. This suggests that creating initial (pure) states with large spatial extent, such as quasi-continuous atom lasers \cite{Johnsson:2007b,Robins:2013}, could yield substantial sensitivity gains. Additionally, compact and/or high-bandwidth devices could benefit from optimal measurements, since shorter interrogation times increase $\Var(G_0(T))$ relative to $F_Q^\text{sc}$.  

For KC interferometers with thermal (mixed) states, Eq.~(\ref{FQ_full}) is only an upper bound for the QFI \cite{Demkowicz-Dobrzanski:2014}. A calculation of $F_Q$ and $F_C$ for thermal sources gives values substantially greater than $F_Q^\text{sc}$ \cite{Stuart_Simon_in_prep}, in qualitative agreement with our above analysis, showing that current thermal-atom gravimetry is suboptimal. However, the QFI and CFI are also smaller than \eq{FQ_full} for thermal sources, suggesting that BECs possess metrological potential beyond what is possible with thermal sources.

Our approach to evaluating matterwave interferometry could significantly influence the design of future state-of-the-art gravimeters. Typical interferometer design assumes a particular form for the measurement signal (e.g., the population difference at the output varies sinusoidally with $g$) and looks no further if there is agreement with simple `best case' formulae such as Eq.~(\ref{delta_g_KC}). In contrast, a Fisher analysis gives the full metrological potential of \emph{any} given dynamical scheme without enforcing such \emph{a priori} assumptions by simply considering the available data. Our matterwave gravimetry analysis opens up new routes to improved sensitivity -- beyond those few implied by Eq.~(\ref{delta_g_KC}). This includes engineering states with high QFI [i.e.~large $\Var(G_0(T))$] and improving information extraction at the interferometer output. Our mirrorless scheme gives a substantial sensitivity boost if high-resolution momentum measurements are available. For \cite{Hardman:2016b}, this momentum resolution is $10^{-4} \hbar k_0$, achievable by further developing the $2\times 10^{-4} \hbar k_0$ resolution measurement of \cite{Gotlibovych:2014}. A Fisher analysis could prove beneficial for evaluating other atom-interferometer-based sensors which produce a complicated output signal, such as schemes utilizing Kapitza-Dirac scattering \cite{Gerlich:2007, Sapiro:2009, Gadway:2009, Li:2014, He:2017, Fekete:2017} or propagation in crossed waveguides \cite{Guarrera:2017}.
  
\begin{acknowledgements}
We acknowledge fruitful discussions with John Close, Chris Freier, Paul Griffin, Kyle Hardman, Guglielmo Tino, Nicola Poli, Samuel Nolan, Augusto Smerzi, and Nicholas Robins. MK and JAD received funding from UK EPSRC through the Networked Quantum Information Technology (NQIT) Hub, grant reference EP/M013243/1. SSS received funding from an Australia Awards-Endeavour Research Fellowship, the Dodd-Walls Centre for Photonic and Quantum Technologies, and Australian Research Council projects DP160104965 and DP150100356. SAH was supported by the European Union's Horizon 2020 research and innovation programme under the Marie Sklodowska-Curie grant agreement No.~704672.  
\end{acknowledgements}

\appendix

\section{QFI of a particle in a gravitational field} \label{appendix_QFI_particle}
Here we give a more detailed derivation of Eq.~(\ref{FQ_grav}). Approximating the gravitational field as a linear potential $m g \hat{z}$, the state of the particle after time $T$ is $\kPT=\hat{U}_g(T)|\Psi_0\rangle$, where
\begin{equation}
\hat{U}_g(T)=\exp\left[-\frac{iT}{\hbar}\left(\frac{\hat{\textbf{p}}^2}{2m}+mg\hz\right)\right].
\end{equation}
In order to isolate the contribution due to the gravitational field $g$, we make use of the Baker-Campbell-Hausdorff (BCH) lemma:
\begin{equation}
e^{\hat{X}+\hat{Y}}=e^{\hat{X}}e^{\hat{Y}}e^{-\frac{1}{2}\left[\hat{X},\hat{Y}\right]}e^{\frac{1}{6}\left(2\left[\hat{Y},[\hat{X},\hat{Y}]\right]+\left[\hat{X},[\hat{X},\hat{Y}]\right]\right)},\label{BCH}
\end{equation}
where $\hat{X}$ and $\hat{Y}$ are operators satisfying the commutation relations
\begin{equation}
	[[[\hat{X},\hat{Y}], \hat{X} ], \hat{X}]=[[[\hat{X},\hat{Y}], \hat{X} ],\hat{Y} ]=[[[\hat{X},\hat{Y}],\hat{Y} ],\hat{Y}]=0.
\end{equation}
This is true for $\hat{X}=-\frac{iT}{\hbar}\frac{\hat{\textbf{p}}^2}{2m}$ and $\hat{Y}=-\frac{iT}{\hbar}mg\hz$, where
\begin{subequations}
\begin{align}
\big[\hat{X},\hat{Y}\big]	&=\frac{i g T^2}{\hbar}\hpz, \\
\Big[\hat{Y},\big[\hat{X},\hat{Y}\big]\Big]	&=\frac{img^2T^3}{\hbar}, \\
\Big[\hat{X},\big[\hat{X},\hat{Y}\big]\Big]	&=0.
\end{align}
\end{subequations}
Thus, \eq{BCH} gives:
\begin{equation}
	e^{-\frac{iT}{\hbar}(\frac{\hat{\textbf{p}}^2}{2m}+mg\hz)}=e^{-\frac{iT}{\hbar}\frac{\hat{\textbf{p}}^2}{2m}} e^{-\frac{iT}{\hbar}mg\hz}e^{-\frac{igT^2}{2\hbar}\hpz}e^{\frac{img^2T^3}{3\hbar}}.
\end{equation}
We use \eq{BCH} again with the choice $\hat{X}=\frac{-iT}{\hbar}mg\hz$ and $\hat{Y}=-\frac{igT^2}{2\hbar}\hpz$, where $\big[\hat{X},\hat{Y}\big]=-\frac{img^2T^3}{2\hbar}$, which allows us to combine $\exp[-i(T/\hbar) mg\hz]$ and $\exp[- igT^2 \hpz / (2\hbar)]$ into a single exponential:
\begin{equation}
	e^{-\frac{iT}{\hbar}mg\hz}e^{-\frac{i g T^2}{2\hbar}\hpz}=e^{-ig\hat{G}_0(T)}e^{-\frac{img^2T^3}{4\hbar}},
\end{equation}
where $\hG(T)=\frac{T}{\hbar}\left(\frac{T}{2}\hpz+m\hz\right)$. Thus, the evolution operator $\hat{U}_g(T)$ can be written as:
\begin{align}
	\hat{U}_g(T)	&=e^{-\frac{iT}{\hbar}\left(\frac{\hat{\textbf{p}}^2}{2m}+mg\hz\right)} \notag \\
				&=\ekto\eg e^{i\frac{mg^2T^3}{12\hbar}}.
\label{ev_op}
\end{align}
We can ignore $\exp[ img^2T^3 / (12\hbar)]$, since this is just a global phase factor, and so the state of the particle after time $T$ is
\begin{equation}
\kPT=e^{-\frac{iT}{\hbar}\frac{\hat{\textbf{p}}^2}{2m}}e^{-ig\hat{G}_0(T)} |\Psi_0\rangle.
\end{equation}
It is now simple to compute the derivative of $\kPT$ with respect to $g$:
\begin{equation}
\kdgP=-ie^{-\frac{iT}{\hbar}\frac{\hat{\textbf{p}}^2}{2m}}\hG(T) e^{-ig\hG(T)}\kPn.
\end{equation}
Consequently,
\begin{subequations}
 \begin{align}
\dgPdgP	&= \langle \Psi_0| \hG(T)^2 | \Psi_0 \rangle, \\
\bPdgP	&=-i\langle \Psi_0| \hG(T) | \Psi_0 \rangle.
\end{align}
\end{subequations}
Substituting these into Eq.~(\ref{general_QFI_equation}) gives our final expression for the QFI, Eq.~(\ref{FQ_grav}).

\section{QFI of a particle after KC interferometry} \label{appendix_QFI_KC}
Here we provide a derivation of Eq.~(\ref{FQKC}). The total evolution of a particle due to KC interferometry is given by the unitary operator
\begin{equation}
	\hat{U}_\text{KC} = \hat{U}_{\frac{\pi}{2}}^{\phi_3}\hat{U}_g(T_2)\hat{U}_\pi^{\phi_2}\hat{U}_g(T_1)\hat{U}_{\frac{\pi}{2}}^{\phi_1},
\end{equation}
where $\hat{U}_{\frac{\pi}{2}}^\phi$ and $\hat{U}_\pi^\phi$ denote $\pi/2$ (50/50 beam splitting) and $\pi$ (mirror) pulses, respectively, and the evolution due to the gravitational field, $\hat{U}_g(T)$, was derived above [see Eq.~(\ref{ev_op})]. This assumes that the $\pi/2$ and $\pi$ pulses are instantaneous (strictly, occur on times much shorter than the interrogation times $T_1$ and $T_2$). 

To begin, the final $\pi/2$ pulse does not change the QFI, whilst the first $\pi/2$ pulse simply gives a new initial state for the particle [see Eq.~(\ref{BS_unitary})]:
\begin{align}
	|\Psi^\prime_0\rangle 	&= \hat{U}_{\frac{\pi}{2}}^{\phi_1}|\Psi_0\rangle \notag \\
						&= \frac{1}{\sqrt{2}}\left(|\psi_0\rangle |a\rangle - i e^{i( k_0 \hat{z}-\phi_1)} |\psi_0\rangle |b\rangle\right), \label{eq_psi_prime}
\end{align}
where $|\Psi_0 \rangle = |a\rangle |\psi_0 \rangle$ and $\phi_1$ is the phase of this first laser pulse. Consequently, the QFI can be computed from the product of operators $\hat{U}_g(T_2)\hat{U}_\pi^{\phi_2}\hat{U}_g(T_1)$, provided expectations are taken with respect to the state $|\Psi'_0\rangle$.

As in Appendix~\ref{appendix_QFI_particle}, our goal is to isolate the $g$-dependence of the evolution. We first consider the product $\hat{U}_g(T_2)\hat{U}_\pi^{\phi_2}$, where [see Eq.~(\ref{BS_unitary})] 
\begin{equation}
	\hup^{\phi_2}=-i\left(e^{-i(k_0\hz-\phi_2)}\ka\bb + e^{i(k_0\hz-\phi_2)}\kb\ba\right), 
\label{up}
\end{equation}
and $\phi_2$ is the phase of this mirror pulse. The BCH lemma \eq{BCH} implies that
\begin{multline}
	e^{\hat{X}} e^{\hat{Y}} e^{-\frac{1}{2}[\hat{X},\hat{Y}]} e^{\frac{1}{6}\left(2\left[\hat{Y},[\hat{X},\hat{Y}]\right]+\left[\hat{X},[\hat{X},\hat{Y}]\right]\right)} \\ = e^{\hat{Y}} e^{\hat{X}} e^{\frac{1}{2}[\hat{X},\hat{Y}]} e^{-\frac{1}{6}\left(2\left[\hat{X},[\hat{X},\hat{Y}]\right]+\left[\hat{Y},[\hat{X},\hat{Y}]\right]\right)}. \label{eq_commute_exp}
\end{multline}
The application of Eq.~(\ref{eq_commute_exp}) with $\hat{X}=-ig\hG(T_2)$ and $\hat{Y}_{\pm}=\pm ik_0\hat{z}$ gives
\begin{align}
	e^{-ig\hG(T_2)}e^{\pm ik_0\hat{z}}=e^{\pm ik_0\hat{z}}e^{-ig\hG(T_2)}e^{\mp i \frac{1}{2} g k_0 T_2^2},
\end{align}
where we have used $\big[\hat{X},\hat{Y}_{\pm}\big]=\mp i g k_0T_2^2 / 2 $. Therefore, after neglecting the global phase factor $\exp[ img^2T_2^3 / (12\hbar)]$ in $\hat{U}_g(T)$:
\begin{widetext}
\begin{align}
	\hat{U}_g(T_2)\hat{U}_\pi^{\phi_2}	&= -i e^{-\frac{iT_2}{\hbar}\frac{\hat{\textbf{p}}^2}{2m}} \left( e^{-i(k_0 \hat{z} - \phi_2)}e^{-ig\hG(T_2)} e^{ i \frac{1}{2} g k_0 T_2^2} | a \rangle \langle b | + e^{i(k_0 \hat{z} + \phi_2)}e^{-ig\hG(T_2)} e^{ -i \frac{1}{2} g k_0 T_2^2} | b \rangle \langle a |\right) \notag \\
				&= -i e^{-\frac{iT_2}{\hbar}\frac{\hat{\textbf{p}}^2}{2m}} \left( e^{-i(k_0 \hat{z} - \phi_2)} e^{ i \frac{1}{2} g k_0 T_2^2} | a \rangle \langle b | + e^{i(k_0 \hat{z} + \phi_2)} e^{ -i \frac{1}{2} g k_0 T_2^2} | b \rangle \langle a |\right) e^{-ig\hG(T_2)}. 
\end{align}
\end{widetext}
Note that $\hG(T_2)$ acts only on the motional state of the particle and therefore commutes with any operators that act on the internal states $|a\rangle$ and $|b\rangle$.

Now internal states $\ka$ and $\kb$ are the eigenvectors of $\hat{S}_z = \frac{1}{2} \left(|a\rangle\langle a| - |b\rangle\langle b|\right)$ satisfying $\hsz\ka=\frac{1}{2}\ka$ and $\hsz\kb=-\frac{1}{2}\kb$. Therefore, for an arbitrary operator $\hat{O}$ which solely acts on the motional state of the particle:
\begin{equation}
	e^{\hat{O}\hsz}\ka =e^{\frac{1}{2}\hat{O}}\ka, \qquad e^{\hat{O}\hsz}\kb =e^{-\frac{1}{2}\hat{O}}\kb. \label{eig_values}
\end{equation}
%
This allows us to write
\begin{subequations}
\begin{align}
	\langle a | e^{ -i \frac{1}{2} g k_0 T_2^2}	&= \langle a | e^{ -i g k_0 T_2^2 \hat{S}_z} = \langle a | e^{ -i g \hat{G}_e}, \\
	\langle b | e^{ i \frac{1}{2} g k_0 T_2^2}	&= \langle b | e^{ -i g k_0 T_2^2 \hat{S}_z} = \langle b | e^{ -i g \hat{G}_e},
\end{align}
\end{subequations}
where $\hat{G}_e=k_0T_2^2\hat{S}_z$. Therefore,
\begin{equation}
	\hat{U}_g(T_2)\hat{U}_\pi^{\phi_2} = e^{-\frac{iT_2}{\hbar}\frac{\hat{\textbf{p}}^2}{2m}}  \hat{U}_\pi^{\phi_2} e^{ -i g \hat{G}_e} e^{-ig\hG(T_2)} .
\end{equation}

Next, we again use Eq.~(\ref{eq_commute_exp}) with $\hat{X}=-ig\gt$ and $\hat{Y}=-\frac{i T_1}{\hb}\frac{\hat{\textbf{p}}^2}{2m}$, where
\begin{subequations}
\begin{align}
\big[\hat{X},\hat{Y}\big]	&=-\frac{igT_1T_2}{\hbar}\hpz, \\
\Big[\hat{X},\big[\hat{X},\hat{Y}\big]\Big]	&=-\frac{img^2T_1T_2^2}{\hbar},
\end{align}
\end{subequations}
to obtain
\begin{multline}
	\egt \eko \\ =\eko\egt e^{-i g \frac{T_1T_2}{\hb}\hat{p}_z}e^{i \frac{m}{2\hb} g^2 T_1T_2^2},
\end{multline}
and therefore (ignoring the global phase factor $\exp[i m g^2 T_1T_2^2 / (2\hb)]$)
\begin{align}
	\hat{U}_g(T_2)\hat{U}_\pi^{\phi_2}\hat{U}_g(T_1) 	&=\ekt\hup^{\phi_2}\ege\eko\egt \notag \\
											& \times e^{-i g \frac{T_1T_2}{\hb} \hat{p}_z}\ego.
\end{align}
We combine the final three exponentials into one using \eq{BCH}:
\begin{multline}
	\hat{U}_g(T_2)\hat{U}_\pi^{\phi_2}\hat{U}_g(T_1)	\\ =\ekt\hup^{\phi_2}\eko e^{-ig\left(\hat{G}_0(T)+\hat{G}_e\right)},
\end{multline}
where $T = T_1 + T_2$ and we have neglected all the global phases produced during the calculation. 

Including the first and second $\pi/2$ pulses (although the second pulse is \emph{not} needed for calculating the QFI), we arrive at the following simplified expression for the full KC interferometer evolution:
\begin{equation}
	\hat{U}_\text{KC}=\hat{U}_0e^{-ig\left(\hat{G}_0(T)+\hat{G}_e\right)}\hat{U}_{\frac{\pi}{2}}^{\phi_1}, \label{eq_U_KC}
\end{equation}
where $\hat{U}_0=\hat{U}_{\frac{\pi}{2}}^{\phi_3}\ekt\hup^{\phi_2}\eko$ is independent of $g$. The state of the particle after interrogation time $T$ is therefore
\begin{equation}
	\kPT=\hat{U}_\text{KC}\kPn =\hat{U}_0e^{-ig\left(\hat{G}_0(T)+\hat{G}_e\right)}|\Psi_0'\rangle,
\end{equation}
which is Eq.~(\ref{psiT_KC_simp}). Taking the derivative with respect to $g$ gives
\begin{subequations}
\begin{align}
\langle\partial_g\Psi(T)|\partial_g\Psi(T)\rangle	&=\langle\Psi_0'|(\hat{G}_0(T)+\hat{G}_e)^2|\Psi_0'\rangle, \\
\langle\Psi(T)|\partial_g\Psi(T)\rangle	&=-i\langle\Psi_0'|( \hat{G}_0(T)+\hat{G}_e)|\Psi_0'\rangle.
\end{align}
\end{subequations}
The QFI is therefore
\begin{equation}
F_Q^\text{KC}=4\mathrm{Var}\left(\hat{G}_0(T)+\hat{G}_e\right),
\end{equation}
where the variance is taken with respect to $|\Psi_0'\rangle$. We use Eq.~(\ref{eq_psi_prime}) to relate this to expectations taken with respect to the initial state $|\Psi_0\rangle$:
\begin{equation}
	F_Q^\text{KC} = 4 \mathrm{Var}(\hat{G}_0(T))  + \frac{1}{4}k_0^2\left(T^2 - 2T_2^2\right)^2,
\end{equation}
which is Eq.~(\ref{FQKC}).

\section{$F_C(\hat{S}_z)$ of KC interferometer} \label{appendix_CFI_KC}
To calculate the CFI $F_C(\hat{S}_z)$ [Eq.~(\ref{FC_anal})], we need to determine expressions for the probabilities $P_a(T)$ and $P_b(T)$ that the particle is detected in state $|a\rangle$ and $|b \rangle$, respectively, at the interferometer output. This first requires expressing $\hat{U}_\text{KC}$ in a more convenient form.
%
To begin, we use Eq.~(\ref{eq_commute_exp}) with $\hat{X}=-i\frac{T_2}{\hbar}\frac{\hat{\textbf{p}}^2}{2m}$ and $\hat{Y}_{\pm}=\pm ik_0\hat{z}$ to obtain
\begin{equation}
	\ekt e^{\pm ik_0\hat{z}}=e^{\pm ik_0\hat{z}}\ekt e^{\mp i\frac{k_0T_2}{m}\hat{p}_z}e^{-i\frac{\hb k_0^2T_2}{2m}},
\end{equation}
where we used $[\hat{X},\hat{Y}_{\pm}]=\mp\frac{ik_0T_2}{m}\hpz$ and $[\hat{Y}_{\pm},[\hat{X},\hat{Y}_{\pm}]]=\frac{i\hbar k_0^2T_2}{m}$. This allows us to commute $\ekt$ and $\hup^{\phi_2}$:
\begin{equation}
\ekt\hup^{\phi_2}=\hup^{\phi_2}\ekt e^{-2i\frac{k_0T_2}{m}\hat{p}_z\hsz}e^{-i\frac{\hb k_0^2T_2}{2m}},
\end{equation}
where we have again used \eq{eig_values}. Neglecting the global phase factor $\exp[-i \hb k_0^2T_2/(2m)]$, we can therefore write Eq.~(\ref{eq_U_KC}) in the convenient form
\begin{equation}
	\hat{U}_\text{KC} = \hat{U}_\text{int}\hat{U}_\text{ext}\hub^{\phi_1},
\end{equation}
where
\begin{align}
	\hat{U}_\text{int}	&\equiv \hub^{\phi_3}\hup^{\phi_2} e^{-2i\frac{k_0T_2}{m}\hat{p}_z\hsz}\ege, \\
	\hat{U}_\text{ext}	&\equiv\ekto\eg.
\end{align}
$\hat{U}_\text{ext}$ only acts on the external (i.e. motional) degrees of freedom, whereas $\hat{U}_\text{int}$ acts on \emph{both} the internal and motional degrees of freedom. Note that $\hui$ and $\hue$ do not commute. 

The state of the particle at the output of the interferometer after interrogation time $T$ is therefore
\begin{align}
	\kPT	&=\hui\hue\hub^{\phi_1}\kPn \notag \\
		&=\frac{1}{\sqrt{2}}\left(\hui\ka\hue\kpn-i\hui\kb\hue e^{i(k_0\hat{z}-\phi_1)}\kpn\right)\label{kPT1}.
\end{align}
From Eq.~(\ref{BS_unitary}) we get:
\begin{multline}
	\hub^{\phi_3}\hup^{\phi_2}	 =-\frac{1}{\sqrt{2}}\left(e^{-i(\phi_2-\phi_3)}\ka\ba+e^{i(\phi_2-\phi_3)}\kb\bb\right) \\
						-\frac{i}{\sqrt{2}}\left(e^{-i(k_0\hat{z}-\phi_2)}\ka\bb+e^{i(k_0\hat{z}-\phi_2)}\kb\ba\right),
\end{multline}
where $\phi_2$ and $\phi_3$ are the phases of the second and the third laser pulses, respectively. Using this and \eq{eig_values}, we obtain
\begin{subequations}
\label{eq_Uint_actions}
\begin{align}
	\hui\ka	&=-\frac{1}{\sqrt{2}}\left[e^{-i(\phi_2-\phi_3)}\ka+ie^{i(k_0\hat{z}-\phi_2)}\kb\right] \notag \\
			& \times e^{-i\frac{k_0T_2}{m}\hat{p}_z}e^{-ig\frac{k_0T_2^2}{2}}, \\
	\hui\kb	&=-\frac{1}{\sqrt{2}}\left[e^{i(\phi_2-\phi_3)}\kb+ie^{-i(k_0\hat{z}-\phi_2)}\ka\right] \notag \\
			& \times e^{i\frac{k_0T_2}{m}\hat{p}_z}e^{ig\frac{k_0T_2^2}{2}}.
\end{align}
\end{subequations}
Substituting Eqs.~(\ref{eq_Uint_actions}) into \eq{kPT1} gives
\begin{widetext}
\begin{align}
	\kPT=-\frac{1}{2}\bigg[\left(e^{-i(\phi_2-\phi_3)}e^{-i\frac{k_0T_2}{m}\hat{p}_z}e^{-ig \frac{k_0T_2^2}{2}}\hue\kpn+e^{-i(k_0\hz-\phi_2)}e^{i\frac{k_0T_2}{m}\hpz}e^{\frac{i}{2}g k_0T_2^2}\hue e^{i(k_0\hz-\phi_1)}\kpn\right)\ka  \nonumber \\
	+i\left(e^{i(k_0\hz-\phi_2)}e^{-i\frac{k_0T_2}{m}\hpz}e^{-ig\frac{k_0T_2^2}{2}}\hue\kpn-e^{i(\phi_2-\phi_3)}e^{i\frac{k_0T_2}{m}\hpz}e^{\frac{i}{2}gk_0T_2^2}\hue e^{i(k_0\hz-\phi_1)}\kpn\right)\kb\bigg].
\end{align}
\end{widetext}

Defining $\kPaT \equiv \langle a\kPT$, the probability of finding the particle in the internal state $\ka$ at the output port of the interferometer is
\begin{align}
	P_a(T) 	&= \langle\Psi_a(T)\kPaT \notag \\
			&= \tfrac{1}{2}[1+\tfrac{1}{2}(e^{i (g k_0T_2^2 - \Phi)}\bpn \hat{Q}\kpn+\mathrm{h.c})], 
\end{align}
where $\Phi \equiv \phi_1-2\phi_2+\phi_3$ and
\begin{align}
	\hat{Q}	&\equiv\egd e^{i\frac{T}{\hb}\frac{\hat{\textbf{p}}^2}{2m}} e^{i\frac{k_0T_2}{m}\hpz}e^{-ik_0\hz}  \notag \\
			&\times e^{i\frac{k_0T_2}{m}\hpz}\ekto\eg e^{ik_0\hz}, \notag \\
			&= e^{i\frac{\hb k_0^2}{2m}(T_2-T_1)}e^{-ig k_0T(T_2-T_1)}e^{-ig \frac{k_0T^2}{2}}e^{i\frac{k_0}{m}(T_2-T_1)\hpz}.
\end{align}
This final simplification follows from repeated application of Eq.~(\ref{eq_commute_exp}), and allows us to express the probability as
\begin{align}
	P_a(T)	&=\tfrac{1}{2}\Big[1+\tfrac{1}{2}\Big(e^{-i \Phi}e^{i\frac{\hb k_0^2}{2m}(T_2-T_1)}e^{-i g k_0(\frac{T^2}{2}-T_1^2)} \notag \\ 
			&\qquad \times \bpn e^{i\frac{k_0}{m}(T_2-T_1)\hpz}\kpn +\mathrm{h.c}\Big)\Big].
\end{align}
If we choose the phases of our laser pulses such that $\phi_1=\phi_2=0$, $\phi_3=\pi / 2$, thereby operating at the point of maximum sensitivity, we can express the probabilities in the following way:
\begin{subequations}
\label{prob}
\begin{align}
P_a(T)	&=\tfrac{1}{2}\left[1 -\tfrac{i}{2}\left(\mathcal{C}e^{i(\phi_f-\phi_g)} - \mathcal{C}^*e^{-i(\phi_f-\phi_g)}\right)\right], \\
P_b(T)	&=\tfrac{1}{2}\left[1 +\tfrac{i}{2}\left(\mathcal{C}e^{i(\phi_f-\phi_g)} - \mathcal{C}^*e^{-i(\phi_f-\phi_g)}\right)\right],
\end{align}
\end{subequations}
where 
\begin{subequations}
\begin{align}
\phi_f	&\equiv\frac{\hbar k_0^2}{2m}(T_2-T_1), \\
\phi_g	&\equiv k_0 g \left(\frac{T^2}{2}-T_1^2\right), \\
\mathcal{C}	&\equiv\langle\psi_0|e^{i\frac{k_0}{m}(T_2-T_1)\hat{p}_z}|\psi_0\rangle.
\end{align}
\end{subequations}
$\phi_f$ represents the phase difference due to the non-symmetrical free evolution of the wavepackets in the two arms of the interferometer, while $\phi_g$ is the phase difference due to gravity. Expressing $\mathcal{C}=|\mathcal{C}|e^{i\vartheta}$ allows us to write Eq.~(\ref{prob}) in the simplified form of Eqs.~(\ref{CFI_probs_equations}).
%
%
Here $|\mathcal{C}|$ is interpreted as a \emph{fringe contrast} and $\alpha=\phi_f-\phi_g+\vartheta$ denotes the total phase shift. 

If we measure the population difference of the two internal states, $\hat{S}_z$, at the output of the interferometer, the CFI is given by
\begin{equation}
F_C(\hsz)=\sum_{j=a,b}\frac{\left(\partial_gP_j\right)^2}{P_j}=\frac{\left(\partial_gP_a\right)^2}{P_aP_b},
\end{equation}
where the last equality follows from the relation $P_a+P_b=1\Rightarrow\partial_gP_a=-\partial_gP_b$. Noting that 
\begin{subequations}
\begin{align}
	P_a(T)P_b(T)	&=\frac{1}{4}\left(1-|\mathcal{C}|^2\sin^2\alpha\right), \\
	\partial_gP_a(T)	&=-\frac{1}{2}|\mathcal{C}| k_0\left(\frac{T^2}{2}-T_1^2\right)\cos\alpha,
\end{align}
\end{subequations}
we arrive at Eq.~(\ref{FC_anal}).

\section{Beam splitter transformation: Derivation of Eq.~(\ref{BS_unitary}).} \label{appendix_BS_derivation}
A Raman beam splitter is typically modelled by the Hamiltonian 
\begin{equation}
\hat{H}_\text{BS} = \frac{\hat{\mathbf{p}}^2}{2m} - \hbar \delta |b\rangle\langle b| +\frac{\hbar \Omega}{2}(|b\rangle\langle a| e^{i(k_0 \hat{z}-\phi)} + \mathrm{h.c.}), \label{H_bs}
\end{equation}
where $\delta$ is the two-photon detuning and $\Omega = \Omega_1\Omega_2/\Delta$ is the effective two-photon Rabi frequency, which depends on the single-photon Rabi frequencies $\Omega_{1,2}$ and the single-photon detuning $\Delta$ \cite{Robins:2006, Johnsson:2007a}. The two-photon detuning is typically set to the two-photon resonance condition $\delta = \hbar k_0^2 / (2m)$. Evolution under this Hamiltonian for a duration $\Delta t$ is given by the unitary time-evolution operator 
\begin{align}
U_\theta^\phi &= \exp\left[\frac{-i \Delta t}{\hbar}\hat{H}_{BS}\right] \nonumber \\
&=  e^{-i\big(\frac{\hat{\mathbf{p}}^2}{2m\hbar} - \frac{\hbar k_0^2}{2m} |b\rangle\langle b|  \big)\frac{\theta}{\Omega} - i\frac{\theta}{2}\big(|b\rangle\langle a| e^{i(k_0 \hat{z}-\phi)} + \mathrm{h.c.}\big)},
\end{align}
where we have defined $\theta = \Omega \Delta t$. If $\hbar \Omega$ is significantly greater than the spread in kinetic energy of the initial state, we can ignore the first term and obtain
\begin{eqnarray}
U_\theta^\phi &=& \exp\left[  -i\frac{\theta}{2}\left(|b\rangle\langle a| e^{i(k_0 \hat{z}-\phi)} + \mathrm{h.c.}\right) \right] \nonumber \\
&=& \hat{1}\cos\left(\tfrac{\theta}{2}\right) - i(|b\rangle\langle a| e^{i(k_0 \hat{z}-\phi)} + \mathrm{h.c.}) \sin\left(\tfrac{\theta}{2}\right), \label{Ubs}
\end{eqnarray}
which is Eq.~(\ref{BS_unitary}). 

\begin{figure}[h]
	\begin{center}
		\includegraphics[width=\columnwidth]{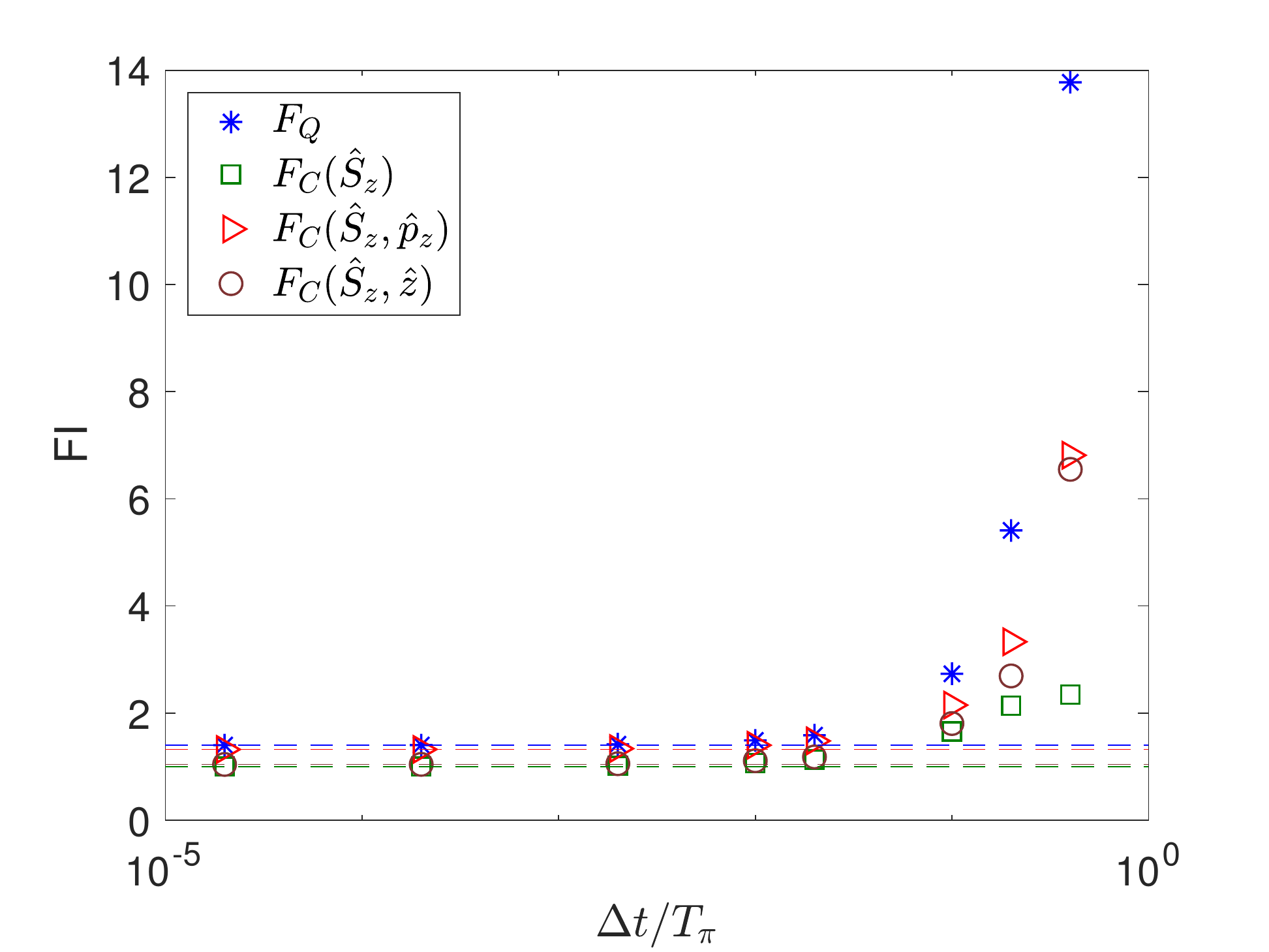}
		\caption{(a) QFI and CFI computed using \eq{H_bs} rather than \eq{Ubs} as a function of $\Delta t$. Provided $\Delta t/T_\pi \ll 1$, \eq{Ubs} (shown by dashed lines of the appropriate colour) is an excellent approximation to the true dynamics. Fisher information is presented in units of $k_0^2 T_\pi^4$.}
		\label{bsfig}
	\end{center}
\end{figure}

Figure~\ref{bsfig} shows the QFI and CFI when the evolution due to the beam splitter and mirror pulses is treated as Schr\"odinger evolution under Hamiltonian~\eq{H_bs}. This evolution was solved numerically for different values of $\Delta t$. We used the same initial state as Fig.~2(a). We set $\Omega$ such that $\Omega \Delta t = \pi/2$ for the two beam splitter pulses, and the duration of the interaction was doubled for the mirror pulse, resulting in $\Omega (2 \Delta t) = \pi$. We find excellent agreement with the ideal beam splitter case as long as $\Delta t \ll T_\pi$. In the regime $\Delta t \sim T_\pi$, there is significant motional dynamics during the beam splitter period, and our approximation is no longer valid. For example, for the maximum value of $\Delta t$ simulated ($\Delta t = 0.4 T_\pi$), the total interferometer sequence time, which is the time from the commencement of the first beam splitter to the conclusion of the second beam splitter, is $3.6 T_\pi$ (compared to $2 T_\pi$ for instantaneous beam splitters). For typical experiments, such as Ref.~\cite{Hardman:2016b}, $\Delta t/T_\pi \sim 10^{-4}$. 

\bibliography{gravity_bib_new}

\begin{thebibliography}{87}%
\makeatletter
\providecommand \@ifxundefined [1]{%
 \@ifx{#1\undefined}
}%
\providecommand \@ifnum [1]{%
 \ifnum #1\expandafter \@firstoftwo
 \else \expandafter \@secondoftwo
 \fi
}%
\providecommand \@ifx [1]{%
 \ifx #1\expandafter \@firstoftwo
 \else \expandafter \@secondoftwo
 \fi
}%
\providecommand \natexlab [1]{#1}%
\providecommand \enquote  [1]{``#1''}%
\providecommand \bibnamefont  [1]{#1}%
\providecommand \bibfnamefont [1]{#1}%
\providecommand \citenamefont [1]{#1}%
\providecommand \href@noop [0]{\@secondoftwo}%
\providecommand \href [0]{\begingroup \@sanitize@url \@href}%
\providecommand \@href[1]{\@@startlink{#1}\@@href}%
\providecommand \@@href[1]{\endgroup#1\@@endlink}%
\providecommand \@sanitize@url [0]{\catcode `\\12\catcode `\$12\catcode
  `\&12\catcode `\#12\catcode `\^12\catcode `\_12\catcode `\%12\relax}%
\providecommand \@@startlink[1]{}%
\providecommand \@@endlink[0]{}%
\providecommand \url  [0]{\begingroup\@sanitize@url \@url }%
\providecommand \@url [1]{\endgroup\@href {#1}{\urlprefix }}%
\providecommand \urlprefix  [0]{URL }%
\providecommand \Eprint [0]{\href }%
\providecommand \doibase [0]{http://dx.doi.org/}%
\providecommand \selectlanguage [0]{\@gobble}%
\providecommand \bibinfo  [0]{\@secondoftwo}%
\providecommand \bibfield  [0]{\@secondoftwo}%
\providecommand \translation [1]{[#1]}%
\providecommand \BibitemOpen [0]{}%
\providecommand \bibitemStop [0]{}%
\providecommand \bibitemNoStop [0]{.\EOS\space}%
\providecommand \EOS [0]{\spacefactor3000\relax}%
\providecommand \BibitemShut  [1]{\csname bibitem#1\endcsname}%
\let\auto@bib@innerbib\@empty
\bibitem [{\citenamefont {Peters}\ \emph {et~al.}(1999)\citenamefont {Peters},
  \citenamefont {Chung},\ and\ \citenamefont {Chu}}]{Peters:1999}%
  \BibitemOpen
  \bibfield  {author} {\bibinfo {author} {\bibfnamefont {Achim}\ \bibnamefont
  {Peters}}, \bibinfo {author} {\bibfnamefont {Keng~Yeow}\ \bibnamefont
  {Chung}}, \ and\ \bibinfo {author} {\bibfnamefont {Steven}\ \bibnamefont
  {Chu}},\ }\bibfield  {title} {\enquote {\bibinfo {title} {Measurement of
  gravitational acceleration by dropping atoms},}\ }\href
  {http://dx.doi.org/10.1038/23655} {\bibfield  {journal} {\bibinfo  {journal}
  {Nature}\ }\textbf {\bibinfo {volume} {400}},\ \bibinfo {pages} {849--852}
  (\bibinfo {year} {1999})}\BibitemShut {NoStop}%
\bibitem [{\citenamefont {Peters}\ \emph {et~al.}(2001)\citenamefont {Peters},
  \citenamefont {Chung},\ and\ \citenamefont {Chu}}]{Peters:2001}%
  \BibitemOpen
  \bibfield  {author} {\bibinfo {author} {\bibfnamefont {A}~\bibnamefont
  {Peters}}, \bibinfo {author} {\bibfnamefont {K~Y}\ \bibnamefont {Chung}}, \
  and\ \bibinfo {author} {\bibfnamefont {S}~\bibnamefont {Chu}},\ }\bibfield
  {title} {\enquote {\bibinfo {title} {High-precision gravity measurements
  using atom interferometry},}\ }\href
  {http://stacks.iop.org/0026-1394/38/i=1/a=4} {\bibfield  {journal} {\bibinfo
  {journal} {Metrologia}\ }\textbf {\bibinfo {volume} {38}},\ \bibinfo {pages}
  {25} (\bibinfo {year} {2001})}\BibitemShut {NoStop}%
\bibitem [{\citenamefont {Hu}\ \emph {et~al.}(2013)\citenamefont {Hu},
  \citenamefont {Sun}, \citenamefont {Duan}, \citenamefont {Zhou},
  \citenamefont {Chen}, \citenamefont {Zhan}, \citenamefont {Zhang},\ and\
  \citenamefont {Luo}}]{Hu:2013}%
  \BibitemOpen
  \bibfield  {author} {\bibinfo {author} {\bibfnamefont {Zhong-Kun}\
  \bibnamefont {Hu}}, \bibinfo {author} {\bibfnamefont {Bu-Liang}\ \bibnamefont
  {Sun}}, \bibinfo {author} {\bibfnamefont {Xiao-Chun}\ \bibnamefont {Duan}},
  \bibinfo {author} {\bibfnamefont {Min-Kang}\ \bibnamefont {Zhou}}, \bibinfo
  {author} {\bibfnamefont {Le-Le}\ \bibnamefont {Chen}}, \bibinfo {author}
  {\bibfnamefont {Su}~\bibnamefont {Zhan}}, \bibinfo {author} {\bibfnamefont
  {Qiao-Zhen}\ \bibnamefont {Zhang}}, \ and\ \bibinfo {author} {\bibfnamefont
  {Jun}\ \bibnamefont {Luo}},\ }\bibfield  {title} {\enquote {\bibinfo {title}
  {Demonstration of an ultrahigh-sensitivity atom-interferometry absolute
  gravimeter},}\ }\href {\doibase 10.1103/PhysRevA.88.043610} {\bibfield
  {journal} {\bibinfo  {journal} {Phys. Rev. A}\ }\textbf {\bibinfo {volume}
  {88}},\ \bibinfo {pages} {043610} (\bibinfo {year} {2013})}\BibitemShut
  {NoStop}%
\bibitem [{\citenamefont {Hauth}\ \emph {et~al.}(2013)\citenamefont {Hauth},
  \citenamefont {Freier}, \citenamefont {Schkolnik}, \citenamefont {Senger},
  \citenamefont {Schmidt},\ and\ \citenamefont {Peters}}]{Hauth:2013}%
  \BibitemOpen
  \bibfield  {author} {\bibinfo {author} {\bibfnamefont {M.}~\bibnamefont
  {Hauth}}, \bibinfo {author} {\bibfnamefont {C.}~\bibnamefont {Freier}},
  \bibinfo {author} {\bibfnamefont {V.}~\bibnamefont {Schkolnik}}, \bibinfo
  {author} {\bibfnamefont {A.}~\bibnamefont {Senger}}, \bibinfo {author}
  {\bibfnamefont {M.}~\bibnamefont {Schmidt}}, \ and\ \bibinfo {author}
  {\bibfnamefont {A.}~\bibnamefont {Peters}},\ }\bibfield  {title} {\enquote
  {\bibinfo {title} {First gravity measurements using the mobile atom
  interferometer gain},}\ }\href {\doibase 10.1007/s00340-013-5413-6}
  {\bibfield  {journal} {\bibinfo  {journal} {Applied Physics B}\ }\textbf
  {\bibinfo {volume} {113}},\ \bibinfo {pages} {49--55} (\bibinfo {year}
  {2013})}\BibitemShut {NoStop}%
\bibitem [{\citenamefont {Altin}\ \emph {et~al.}(2013)\citenamefont {Altin},
  \citenamefont {Johnsson}, \citenamefont {Negnevitsky}, \citenamefont
  {Dennis}, \citenamefont {Anderson}, \citenamefont {Debs}, \citenamefont
  {Szigeti}, \citenamefont {Hardman}, \citenamefont {Bennetts}, \citenamefont
  {McDonald}, \citenamefont {Turner}, \citenamefont {Close},\ and\
  \citenamefont {Robins}}]{Altin:2013}%
  \BibitemOpen
  \bibfield  {author} {\bibinfo {author} {\bibfnamefont {P~A}\ \bibnamefont
  {Altin}}, \bibinfo {author} {\bibfnamefont {M~T}\ \bibnamefont {Johnsson}},
  \bibinfo {author} {\bibfnamefont {V}~\bibnamefont {Negnevitsky}}, \bibinfo
  {author} {\bibfnamefont {G~R}\ \bibnamefont {Dennis}}, \bibinfo {author}
  {\bibfnamefont {R~P}\ \bibnamefont {Anderson}}, \bibinfo {author}
  {\bibfnamefont {J~E}\ \bibnamefont {Debs}}, \bibinfo {author} {\bibfnamefont
  {S~S}\ \bibnamefont {Szigeti}}, \bibinfo {author} {\bibfnamefont {K~S}\
  \bibnamefont {Hardman}}, \bibinfo {author} {\bibfnamefont {S}~\bibnamefont
  {Bennetts}}, \bibinfo {author} {\bibfnamefont {G~D}\ \bibnamefont
  {McDonald}}, \bibinfo {author} {\bibfnamefont {L~D}\ \bibnamefont {Turner}},
  \bibinfo {author} {\bibfnamefont {J~D}\ \bibnamefont {Close}}, \ and\
  \bibinfo {author} {\bibfnamefont {N~P}\ \bibnamefont {Robins}},\ }\bibfield
  {title} {\enquote {\bibinfo {title} {Precision atomic gravimeter based on
  bragg diffraction},}\ }\href
  {http://stacks.iop.org/1367-2630/15/i=2/a=023009} {\bibfield  {journal}
  {\bibinfo  {journal} {New Journal of Physics}\ }\textbf {\bibinfo {volume}
  {15}},\ \bibinfo {pages} {023009} (\bibinfo {year} {2013})}\BibitemShut
  {NoStop}%
\bibitem [{\citenamefont {Freier}\ \emph {et~al.}(2016)\citenamefont {Freier},
  \citenamefont {Hauth}, \citenamefont {Schkolnik}, \citenamefont {Leykauf},
  \citenamefont {Schilling}, \citenamefont {Wziontek}, \citenamefont
  {Scherneck}, \citenamefont {M{\"u}ller},\ and\ \citenamefont
  {Peters}}]{Freier:2016}%
  \BibitemOpen
  \bibfield  {author} {\bibinfo {author} {\bibfnamefont {C}~\bibnamefont
  {Freier}}, \bibinfo {author} {\bibfnamefont {M}~\bibnamefont {Hauth}},
  \bibinfo {author} {\bibfnamefont {V}~\bibnamefont {Schkolnik}}, \bibinfo
  {author} {\bibfnamefont {B}~\bibnamefont {Leykauf}}, \bibinfo {author}
  {\bibfnamefont {M}~\bibnamefont {Schilling}}, \bibinfo {author}
  {\bibfnamefont {H}~\bibnamefont {Wziontek}}, \bibinfo {author} {\bibfnamefont
  {H-G}\ \bibnamefont {Scherneck}}, \bibinfo {author} {\bibfnamefont
  {J}~\bibnamefont {M{\"u}ller}}, \ and\ \bibinfo {author} {\bibfnamefont
  {A}~\bibnamefont {Peters}},\ }\bibfield  {title} {\enquote {\bibinfo {title}
  {Mobile quantum gravity sensor with unprecedented stability},}\ }\href
  {http://stacks.iop.org/1742-6596/723/i=1/a=012050} {\bibfield  {journal}
  {\bibinfo  {journal} {Journal of Physics: Conference Series}\ }\textbf
  {\bibinfo {volume} {723}},\ \bibinfo {pages} {012050} (\bibinfo {year}
  {2016})}\BibitemShut {NoStop}%
\bibitem [{\citenamefont {Hardman}\ \emph
  {et~al.}(2016{\natexlab{a}})\citenamefont {Hardman}, \citenamefont {Everitt},
  \citenamefont {McDonald}, \citenamefont {Manju}, \citenamefont {Wigley},
  \citenamefont {Sooriyabandara}, \citenamefont {Kuhn}, \citenamefont {Debs},
  \citenamefont {Close},\ and\ \citenamefont {Robins}}]{Hardman:2016b}%
  \BibitemOpen
  \bibfield  {author} {\bibinfo {author} {\bibfnamefont {K.~S.}\ \bibnamefont
  {Hardman}}, \bibinfo {author} {\bibfnamefont {P.~J.}\ \bibnamefont
  {Everitt}}, \bibinfo {author} {\bibfnamefont {G.~D.}\ \bibnamefont
  {McDonald}}, \bibinfo {author} {\bibfnamefont {P.}~\bibnamefont {Manju}},
  \bibinfo {author} {\bibfnamefont {P.~B.}\ \bibnamefont {Wigley}}, \bibinfo
  {author} {\bibfnamefont {M.~A.}\ \bibnamefont {Sooriyabandara}}, \bibinfo
  {author} {\bibfnamefont {C.~C.~N.}\ \bibnamefont {Kuhn}}, \bibinfo {author}
  {\bibfnamefont {J.~E.}\ \bibnamefont {Debs}}, \bibinfo {author}
  {\bibfnamefont {J.~D.}\ \bibnamefont {Close}}, \ and\ \bibinfo {author}
  {\bibfnamefont {N.~P.}\ \bibnamefont {Robins}},\ }\bibfield  {title}
  {\enquote {\bibinfo {title} {Simultaneous precision gravimetry and magnetic
  gradiometry with a bose-einstein condensate: A high precision, quantum
  sensor},}\ }\href {\doibase 10.1103/PhysRevLett.117.138501} {\bibfield
  {journal} {\bibinfo  {journal} {Phys. Rev. Lett.}\ }\textbf {\bibinfo
  {volume} {117}},\ \bibinfo {pages} {138501} (\bibinfo {year}
  {2016}{\natexlab{a}})}\BibitemShut {NoStop}%
\bibitem [{\citenamefont {Snadden}\ \emph {et~al.}(1998)\citenamefont
  {Snadden}, \citenamefont {McGuirk}, \citenamefont {Bouyer}, \citenamefont
  {Haritos},\ and\ \citenamefont {Kasevich}}]{Snadden:1998}%
  \BibitemOpen
  \bibfield  {author} {\bibinfo {author} {\bibfnamefont {M.~J.}\ \bibnamefont
  {Snadden}}, \bibinfo {author} {\bibfnamefont {J.~M.}\ \bibnamefont
  {McGuirk}}, \bibinfo {author} {\bibfnamefont {P.}~\bibnamefont {Bouyer}},
  \bibinfo {author} {\bibfnamefont {K.~G.}\ \bibnamefont {Haritos}}, \ and\
  \bibinfo {author} {\bibfnamefont {M.~A.}\ \bibnamefont {Kasevich}},\
  }\bibfield  {title} {\enquote {\bibinfo {title} {Measurement of the earth's
  gravity gradient with an atom interferometer-based gravity gradiometer},}\
  }\href {\doibase 10.1103/PhysRevLett.81.971} {\bibfield  {journal} {\bibinfo
  {journal} {Phys. Rev. Lett.}\ }\textbf {\bibinfo {volume} {81}},\ \bibinfo
  {pages} {971--974} (\bibinfo {year} {1998})}\BibitemShut {NoStop}%
\bibitem [{\citenamefont {Stern}\ \emph {et~al.}(2009)\citenamefont {Stern},
  \citenamefont {Battelier}, \citenamefont {Geiger}, \citenamefont {Varoquaux},
  \citenamefont {Villing}, \citenamefont {Moron}, \citenamefont {Carraz},
  \citenamefont {Zahzam}, \citenamefont {Bidel}, \citenamefont {Chaibi},
  \citenamefont {Pereira Dos~Santos}, \citenamefont {Bresson}, \citenamefont
  {Landragin},\ and\ \citenamefont {Bouyer}}]{Stern:2009}%
  \BibitemOpen
  \bibfield  {author} {\bibinfo {author} {\bibfnamefont {G.}~\bibnamefont
  {Stern}}, \bibinfo {author} {\bibfnamefont {B.}~\bibnamefont {Battelier}},
  \bibinfo {author} {\bibfnamefont {R.}~\bibnamefont {Geiger}}, \bibinfo
  {author} {\bibfnamefont {G.}~\bibnamefont {Varoquaux}}, \bibinfo {author}
  {\bibfnamefont {A.}~\bibnamefont {Villing}}, \bibinfo {author} {\bibfnamefont
  {F.}~\bibnamefont {Moron}}, \bibinfo {author} {\bibfnamefont
  {O.}~\bibnamefont {Carraz}}, \bibinfo {author} {\bibfnamefont
  {N.}~\bibnamefont {Zahzam}}, \bibinfo {author} {\bibfnamefont
  {Y.}~\bibnamefont {Bidel}}, \bibinfo {author} {\bibfnamefont
  {W.}~\bibnamefont {Chaibi}}, \bibinfo {author} {\bibfnamefont
  {F.}~\bibnamefont {Pereira Dos~Santos}}, \bibinfo {author} {\bibfnamefont
  {A.}~\bibnamefont {Bresson}}, \bibinfo {author} {\bibfnamefont
  {A.}~\bibnamefont {Landragin}}, \ and\ \bibinfo {author} {\bibfnamefont
  {P.}~\bibnamefont {Bouyer}},\ }\bibfield  {title} {\enquote {\bibinfo {title}
  {Light-pulse atom interferometry in microgravity},}\ }\href {\doibase
  10.1140/epjd/e2009-00150-5} {\bibfield  {journal} {\bibinfo  {journal} {The
  European Physical Journal D}\ }\textbf {\bibinfo {volume} {53}},\ \bibinfo
  {pages} {353--357} (\bibinfo {year} {2009})}\BibitemShut {NoStop}%
\bibitem [{\citenamefont {Sorrentino}\ \emph {et~al.}(2014)\citenamefont
  {Sorrentino}, \citenamefont {Bodart}, \citenamefont {Cacciapuoti},
  \citenamefont {Lien}, \citenamefont {Prevedelli}, \citenamefont {Rosi},
  \citenamefont {Salvi},\ and\ \citenamefont {Tino}}]{Sorrentino:2014}%
  \BibitemOpen
  \bibfield  {author} {\bibinfo {author} {\bibfnamefont {F.}~\bibnamefont
  {Sorrentino}}, \bibinfo {author} {\bibfnamefont {Q.}~\bibnamefont {Bodart}},
  \bibinfo {author} {\bibfnamefont {L.}~\bibnamefont {Cacciapuoti}}, \bibinfo
  {author} {\bibfnamefont {Y.-H.}\ \bibnamefont {Lien}}, \bibinfo {author}
  {\bibfnamefont {M.}~\bibnamefont {Prevedelli}}, \bibinfo {author}
  {\bibfnamefont {G.}~\bibnamefont {Rosi}}, \bibinfo {author} {\bibfnamefont
  {L.}~\bibnamefont {Salvi}}, \ and\ \bibinfo {author} {\bibfnamefont {G.~M.}\
  \bibnamefont {Tino}},\ }\bibfield  {title} {\enquote {\bibinfo {title}
  {Sensitivity limits of a raman atom interferometer as a gravity
  gradiometer},}\ }\href {\doibase 10.1103/PhysRevA.89.023607} {\bibfield
  {journal} {\bibinfo  {journal} {Phys. Rev. A}\ }\textbf {\bibinfo {volume}
  {89}},\ \bibinfo {pages} {023607} (\bibinfo {year} {2014})}\BibitemShut
  {NoStop}%
\bibitem [{\citenamefont {Biedermann}\ \emph {et~al.}(2015)\citenamefont
  {Biedermann}, \citenamefont {Wu}, \citenamefont {Deslauriers}, \citenamefont
  {Roy}, \citenamefont {Mahadeswaraswamy},\ and\ \citenamefont
  {Kasevich}}]{Biedermann:2015}%
  \BibitemOpen
  \bibfield  {author} {\bibinfo {author} {\bibfnamefont {G.~W.}\ \bibnamefont
  {Biedermann}}, \bibinfo {author} {\bibfnamefont {X.}~\bibnamefont {Wu}},
  \bibinfo {author} {\bibfnamefont {L.}~\bibnamefont {Deslauriers}}, \bibinfo
  {author} {\bibfnamefont {S.}~\bibnamefont {Roy}}, \bibinfo {author}
  {\bibfnamefont {C.}~\bibnamefont {Mahadeswaraswamy}}, \ and\ \bibinfo
  {author} {\bibfnamefont {M.~A.}\ \bibnamefont {Kasevich}},\ }\bibfield
  {title} {\enquote {\bibinfo {title} {Testing gravity with cold-atom
  interferometers},}\ }\href {\doibase 10.1103/PhysRevA.91.033629} {\bibfield
  {journal} {\bibinfo  {journal} {Phys. Rev. A}\ }\textbf {\bibinfo {volume}
  {91}},\ \bibinfo {pages} {033629} (\bibinfo {year} {2015})}\BibitemShut
  {NoStop}%
\bibitem [{\citenamefont {D'Amico}\ \emph {et~al.}(2016)\citenamefont
  {D'Amico}, \citenamefont {Borselli}, \citenamefont {Cacciapuoti},
  \citenamefont {Prevedelli}, \citenamefont {Rosi}, \citenamefont
  {Sorrentino},\ and\ \citenamefont {Tino}}]{DAmico:2016}%
  \BibitemOpen
  \bibfield  {author} {\bibinfo {author} {\bibfnamefont {G.}~\bibnamefont
  {D'Amico}}, \bibinfo {author} {\bibfnamefont {F.}~\bibnamefont {Borselli}},
  \bibinfo {author} {\bibfnamefont {L.}~\bibnamefont {Cacciapuoti}}, \bibinfo
  {author} {\bibfnamefont {M.}~\bibnamefont {Prevedelli}}, \bibinfo {author}
  {\bibfnamefont {G.}~\bibnamefont {Rosi}}, \bibinfo {author} {\bibfnamefont
  {F.}~\bibnamefont {Sorrentino}}, \ and\ \bibinfo {author} {\bibfnamefont
  {G.~M.}\ \bibnamefont {Tino}},\ }\bibfield  {title} {\enquote {\bibinfo
  {title} {Bragg interferometer for gravity gradient measurements},}\ }\href
  {\doibase 10.1103/PhysRevA.93.063628} {\bibfield  {journal} {\bibinfo
  {journal} {Phys. Rev. A}\ }\textbf {\bibinfo {volume} {93}},\ \bibinfo
  {pages} {063628} (\bibinfo {year} {2016})}\BibitemShut {NoStop}%
\bibitem [{\citenamefont {Asenbaum}\ \emph {et~al.}(2017)\citenamefont
  {Asenbaum}, \citenamefont {Overstreet}, \citenamefont {Kovachy},
  \citenamefont {Brown}, \citenamefont {Hogan},\ and\ \citenamefont
  {Kasevich}}]{Asenbaum:2017}%
  \BibitemOpen
  \bibfield  {author} {\bibinfo {author} {\bibfnamefont {Peter}\ \bibnamefont
  {Asenbaum}}, \bibinfo {author} {\bibfnamefont {Chris}\ \bibnamefont
  {Overstreet}}, \bibinfo {author} {\bibfnamefont {Tim}\ \bibnamefont
  {Kovachy}}, \bibinfo {author} {\bibfnamefont {Daniel~D.}\ \bibnamefont
  {Brown}}, \bibinfo {author} {\bibfnamefont {Jason~M.}\ \bibnamefont {Hogan}},
  \ and\ \bibinfo {author} {\bibfnamefont {Mark~A.}\ \bibnamefont {Kasevich}},\
  }\bibfield  {title} {\enquote {\bibinfo {title} {Phase shift in an atom
  interferometer due to spacetime curvature across its wave function},}\ }\href
  {\doibase 10.1103/PhysRevLett.118.183602} {\bibfield  {journal} {\bibinfo
  {journal} {Phys. Rev. Lett.}\ }\textbf {\bibinfo {volume} {118}},\ \bibinfo
  {pages} {183602} (\bibinfo {year} {2017})}\BibitemShut {NoStop}%
\bibitem [{\citenamefont {Bidel}\ \emph {et~al.}(2018)\citenamefont {Bidel},
  \citenamefont {Zahzam}, \citenamefont {Blanchard}, \citenamefont {Bonnin},
  \citenamefont {Cadoret}, \citenamefont {Bresson}, \citenamefont {Rouxel},\
  and\ \citenamefont {Lequentrec-Lalancette}}]{Bidel:2018}%
  \BibitemOpen
  \bibfield  {author} {\bibinfo {author} {\bibfnamefont {Y.}~\bibnamefont
  {Bidel}}, \bibinfo {author} {\bibfnamefont {N.}~\bibnamefont {Zahzam}},
  \bibinfo {author} {\bibfnamefont {C.}~\bibnamefont {Blanchard}}, \bibinfo
  {author} {\bibfnamefont {A.}~\bibnamefont {Bonnin}}, \bibinfo {author}
  {\bibfnamefont {M.}~\bibnamefont {Cadoret}}, \bibinfo {author} {\bibfnamefont
  {A.}~\bibnamefont {Bresson}}, \bibinfo {author} {\bibfnamefont
  {D.}~\bibnamefont {Rouxel}}, \ and\ \bibinfo {author} {\bibfnamefont {M.~F.}\
  \bibnamefont {Lequentrec-Lalancette}},\ }\bibfield  {title} {\enquote
  {\bibinfo {title} {Absolute marine gravimetry with matter-wave
  interferometry},}\ }\href {\doibase 10.1038/s41467-018-03040-2} {\bibfield
  {journal} {\bibinfo  {journal} {Nature Communications}\ }\textbf {\bibinfo
  {volume} {9}},\ \bibinfo {pages} {627} (\bibinfo {year} {2018})}\BibitemShut
  {NoStop}%
\bibitem [{\citenamefont {Richeson}(2008)}]{Richeson:2008}%
  \BibitemOpen
  \bibfield  {author} {\bibinfo {author} {\bibfnamefont {Justin~Arthur}\
  \bibnamefont {Richeson}},\ }\emph {\bibinfo {title} {Gravity Gradiometer
  Aided Inertial Navigation Within Non-{GNSS} Environments}},\ \href@noop {}
  {Ph.D. thesis},\ \bibinfo  {school} {University of Maryland} (\bibinfo {year}
  {2008})\BibitemShut {NoStop}%
\bibitem [{\citenamefont {Evstifeev}(2017)}]{Evstifeev:2017}%
  \BibitemOpen
  \bibfield  {author} {\bibinfo {author} {\bibfnamefont {M.~I.}\ \bibnamefont
  {Evstifeev}},\ }\bibfield  {title} {\enquote {\bibinfo {title} {The state of
  the art in the development of onboard gravity gradiometers},}\ }\href
  {\doibase 10.1134/S2075108717010047} {\bibfield  {journal} {\bibinfo
  {journal} {Gyroscopy and Navigation}\ }\textbf {\bibinfo {volume} {8}},\
  \bibinfo {pages} {68--79} (\bibinfo {year} {2017})}\BibitemShut {NoStop}%
\bibitem [{\citenamefont {Fray}\ \emph {et~al.}(2004)\citenamefont {Fray},
  \citenamefont {Diez}, \citenamefont {H\"ansch},\ and\ \citenamefont
  {Weitz}}]{Fray:2004}%
  \BibitemOpen
  \bibfield  {author} {\bibinfo {author} {\bibfnamefont {Sebastian}\
  \bibnamefont {Fray}}, \bibinfo {author} {\bibfnamefont {Cristina~Alvarez}\
  \bibnamefont {Diez}}, \bibinfo {author} {\bibfnamefont {Theodor~W.}\
  \bibnamefont {H\"ansch}}, \ and\ \bibinfo {author} {\bibfnamefont {Martin}\
  \bibnamefont {Weitz}},\ }\bibfield  {title} {\enquote {\bibinfo {title}
  {Atomic interferometer with amplitude gratings of light and its applications
  to atom based tests of the equivalence principle},}\ }\href {\doibase
  10.1103/PhysRevLett.93.240404} {\bibfield  {journal} {\bibinfo  {journal}
  {Phys. Rev. Lett.}\ }\textbf {\bibinfo {volume} {93}},\ \bibinfo {pages}
  {240404} (\bibinfo {year} {2004})}\BibitemShut {NoStop}%
\bibitem [{\citenamefont {Dimopoulos}\ \emph {et~al.}(2007)\citenamefont
  {Dimopoulos}, \citenamefont {Graham}, \citenamefont {Hogan},\ and\
  \citenamefont {Kasevich}}]{Dimopoulos:2007}%
  \BibitemOpen
  \bibfield  {author} {\bibinfo {author} {\bibfnamefont {Savas}\ \bibnamefont
  {Dimopoulos}}, \bibinfo {author} {\bibfnamefont {Peter~W.}\ \bibnamefont
  {Graham}}, \bibinfo {author} {\bibfnamefont {Jason~M.}\ \bibnamefont
  {Hogan}}, \ and\ \bibinfo {author} {\bibfnamefont {Mark~A.}\ \bibnamefont
  {Kasevich}},\ }\bibfield  {title} {\enquote {\bibinfo {title} {Testing
  general relativity with atom interferometry},}\ }\href {\doibase
  10.1103/PhysRevLett.98.111102} {\bibfield  {journal} {\bibinfo  {journal}
  {Phys. Rev. Lett.}\ }\textbf {\bibinfo {volume} {98}},\ \bibinfo {pages}
  {111102} (\bibinfo {year} {2007})}\BibitemShut {NoStop}%
\bibitem [{\citenamefont {Schlippert}\ \emph {et~al.}(2014)\citenamefont
  {Schlippert}, \citenamefont {Hartwig}, \citenamefont {Albers}, \citenamefont
  {Richardson}, \citenamefont {Schubert}, \citenamefont {Roura}, \citenamefont
  {Schleich}, \citenamefont {Ertmer},\ and\ \citenamefont
  {Rasel}}]{Schlippert:2014}%
  \BibitemOpen
  \bibfield  {author} {\bibinfo {author} {\bibfnamefont {D.}~\bibnamefont
  {Schlippert}}, \bibinfo {author} {\bibfnamefont {J.}~\bibnamefont {Hartwig}},
  \bibinfo {author} {\bibfnamefont {H.}~\bibnamefont {Albers}}, \bibinfo
  {author} {\bibfnamefont {L.~L.}\ \bibnamefont {Richardson}}, \bibinfo
  {author} {\bibfnamefont {C.}~\bibnamefont {Schubert}}, \bibinfo {author}
  {\bibfnamefont {A.}~\bibnamefont {Roura}}, \bibinfo {author} {\bibfnamefont
  {W.~P.}\ \bibnamefont {Schleich}}, \bibinfo {author} {\bibfnamefont
  {W.}~\bibnamefont {Ertmer}}, \ and\ \bibinfo {author} {\bibfnamefont {E.~M.}\
  \bibnamefont {Rasel}},\ }\bibfield  {title} {\enquote {\bibinfo {title}
  {Quantum test of the universality of free fall},}\ }\href {\doibase
  10.1103/PhysRevLett.112.203002} {\bibfield  {journal} {\bibinfo  {journal}
  {Phys. Rev. Lett.}\ }\textbf {\bibinfo {volume} {112}},\ \bibinfo {pages}
  {203002} (\bibinfo {year} {2014})}\BibitemShut {NoStop}%
\bibitem [{\citenamefont {Amelino-Camelia}\ \emph {et~al.}(2009)\citenamefont
  {Amelino-Camelia}, \citenamefont {L\"ammerzahl}, \citenamefont {Mercati},\
  and\ \citenamefont {Tino}}]{Amelino-Camelia:2009}%
  \BibitemOpen
  \bibfield  {author} {\bibinfo {author} {\bibfnamefont {Giovanni}\
  \bibnamefont {Amelino-Camelia}}, \bibinfo {author} {\bibfnamefont {Claus}\
  \bibnamefont {L\"ammerzahl}}, \bibinfo {author} {\bibfnamefont {Flavio}\
  \bibnamefont {Mercati}}, \ and\ \bibinfo {author} {\bibfnamefont
  {Guglielmo~M.}\ \bibnamefont {Tino}},\ }\bibfield  {title} {\enquote
  {\bibinfo {title} {Constraining the energy-momentum dispersion relation with
  planck-scale sensitivity using cold atoms},}\ }\href {\doibase
  10.1103/PhysRevLett.103.171302} {\bibfield  {journal} {\bibinfo  {journal}
  {Phys. Rev. Lett.}\ }\textbf {\bibinfo {volume} {103}},\ \bibinfo {pages}
  {171302} (\bibinfo {year} {2009})}\BibitemShut {NoStop}%
\bibitem [{\citenamefont {Gao}\ and\ \citenamefont {Zhan}(2016)}]{Gao:2016}%
  \BibitemOpen
  \bibfield  {author} {\bibinfo {author} {\bibfnamefont {Dongfeng}\
  \bibnamefont {Gao}}\ and\ \bibinfo {author} {\bibfnamefont {Mingsheng}\
  \bibnamefont {Zhan}},\ }\bibfield  {title} {\enquote {\bibinfo {title}
  {Constraining the generalized uncertainty principle with cold atoms},}\
  }\href {\doibase 10.1103/PhysRevA.94.013607} {\bibfield  {journal} {\bibinfo
  {journal} {Phys. Rev. A}\ }\textbf {\bibinfo {volume} {94}},\ \bibinfo
  {pages} {013607} (\bibinfo {year} {2016})}\BibitemShut {NoStop}%
\bibitem [{\citenamefont {Bord{\'e}}(1989)}]{Borde:1989}%
  \BibitemOpen
  \bibfield  {author} {\bibinfo {author} {\bibfnamefont {Ch.J.}\ \bibnamefont
  {Bord{\'e}}},\ }\bibfield  {title} {\enquote {\bibinfo {title} {Atomic
  interferometry with internal state labelling},}\ }\href@noop {} {\bibfield
  {journal} {\bibinfo  {journal} {Physics Letters A}\ }\textbf {\bibinfo
  {volume} {140}},\ \bibinfo {pages} {10 -- 12} (\bibinfo {year}
  {1989})}\BibitemShut {NoStop}%
\bibitem [{\citenamefont {Kasevich}\ and\ \citenamefont
  {Chu}(1991)}]{Kasevich:1991}%
  \BibitemOpen
  \bibfield  {author} {\bibinfo {author} {\bibfnamefont {Mark}\ \bibnamefont
  {Kasevich}}\ and\ \bibinfo {author} {\bibfnamefont {Steven}\ \bibnamefont
  {Chu}},\ }\bibfield  {title} {\enquote {\bibinfo {title} {Atomic
  interferometry using stimulated raman transitions},}\ }\href {\doibase
  10.1103/PhysRevLett.67.181} {\bibfield  {journal} {\bibinfo  {journal} {Phys.
  Rev. Lett.}\ }\textbf {\bibinfo {volume} {67}},\ \bibinfo {pages} {181--184}
  (\bibinfo {year} {1991})}\BibitemShut {NoStop}%
\bibitem [{\citenamefont {Kasevich}\ and\ \citenamefont
  {Chu}(1992)}]{Kasevich:1992}%
  \BibitemOpen
  \bibfield  {author} {\bibinfo {author} {\bibfnamefont {M.}~\bibnamefont
  {Kasevich}}\ and\ \bibinfo {author} {\bibfnamefont {S.}~\bibnamefont {Chu}},\
  }\bibfield  {title} {\enquote {\bibinfo {title} {Measurement of the
  gravitational acceleration of an atom with a light-pulse atom
  interferometer},}\ }\href {http://dx.doi.org/10.1007/BF00325375} {\bibfield
  {journal} {\bibinfo  {journal} {Applied Physics B: Lasers and Optics}\
  }\textbf {\bibinfo {volume} {54}},\ \bibinfo {pages} {321--332} (\bibinfo
  {year} {1992})}\BibitemShut {NoStop}%
\bibitem [{\citenamefont {Storey}\ and\ \citenamefont
  {Cohen-Tannoudji}(1994)}]{Storey:1994}%
  \BibitemOpen
  \bibfield  {author} {\bibinfo {author} {\bibfnamefont {Pippa}\ \bibnamefont
  {Storey}}\ and\ \bibinfo {author} {\bibfnamefont {Claude}\ \bibnamefont
  {Cohen-Tannoudji}},\ }\bibfield  {title} {\enquote {\bibinfo {title} {The
  feynman path integral approach to atomic interferometry. a tutorial},}\
  }\href@noop {} {\bibfield  {journal} {\bibinfo  {journal} {Journal de
  Physique II}\ }\textbf {\bibinfo {volume} {4}},\ \bibinfo {pages}
  {1999--2027} (\bibinfo {year} {1994})}\BibitemShut {NoStop}%
\bibitem [{\citenamefont {Schleich}\ \emph
  {et~al.}(2013{\natexlab{a}})\citenamefont {Schleich}, \citenamefont
  {Greenberger},\ and\ \citenamefont {Rasel}}]{Schleich:2013}%
  \BibitemOpen
  \bibfield  {author} {\bibinfo {author} {\bibfnamefont {Wolfgang~P.}\
  \bibnamefont {Schleich}}, \bibinfo {author} {\bibfnamefont {Daniel~M.}\
  \bibnamefont {Greenberger}}, \ and\ \bibinfo {author} {\bibfnamefont
  {Ernst~M.}\ \bibnamefont {Rasel}},\ }\bibfield  {title} {\enquote {\bibinfo
  {title} {Redshift controversy in atom interferometry: Representation
  dependence of the origin of phase shift},}\ }\href {\doibase
  10.1103/PhysRevLett.110.010401} {\bibfield  {journal} {\bibinfo  {journal}
  {Phys. Rev. Lett.}\ }\textbf {\bibinfo {volume} {110}},\ \bibinfo {pages}
  {010401} (\bibinfo {year} {2013}{\natexlab{a}})}\BibitemShut {NoStop}%
\bibitem [{\citenamefont {Schleich}\ \emph
  {et~al.}(2013{\natexlab{b}})\citenamefont {Schleich}, \citenamefont
  {Greenberger},\ and\ \citenamefont {Rasel}}]{Schleich:2013b}%
  \BibitemOpen
  \bibfield  {author} {\bibinfo {author} {\bibfnamefont {Wolfgang~P}\
  \bibnamefont {Schleich}}, \bibinfo {author} {\bibfnamefont {Daniel~M}\
  \bibnamefont {Greenberger}}, \ and\ \bibinfo {author} {\bibfnamefont
  {Ernst~M}\ \bibnamefont {Rasel}},\ }\bibfield  {title} {\enquote {\bibinfo
  {title} {A representation-free description of the {K}asevich--{C}hu
  interferometer: a resolution of the redshift controversy},}\ }\href
  {http://stacks.iop.org/1367-2630/15/i=1/a=013007} {\bibfield  {journal}
  {\bibinfo  {journal} {New Journal of Physics}\ }\textbf {\bibinfo {volume}
  {15}},\ \bibinfo {pages} {013007} (\bibinfo {year}
  {2013}{\natexlab{b}})}\BibitemShut {NoStop}%
\bibitem [{\citenamefont {M\"uller}\ \emph {et~al.}(2008)\citenamefont
  {M\"uller}, \citenamefont {Chiow}, \citenamefont {Long}, \citenamefont
  {Herrmann},\ and\ \citenamefont {Chu}}]{Muller:2008b}%
  \BibitemOpen
  \bibfield  {author} {\bibinfo {author} {\bibfnamefont {Holger}\ \bibnamefont
  {M\"uller}}, \bibinfo {author} {\bibfnamefont {Sheng-wey}\ \bibnamefont
  {Chiow}}, \bibinfo {author} {\bibfnamefont {Quan}\ \bibnamefont {Long}},
  \bibinfo {author} {\bibfnamefont {Sven}\ \bibnamefont {Herrmann}}, \ and\
  \bibinfo {author} {\bibfnamefont {Steven}\ \bibnamefont {Chu}},\ }\bibfield
  {title} {\enquote {\bibinfo {title} {Atom interferometry with up to
  24-photon-momentum-transfer beam splitters},}\ }\href {\doibase
  10.1103/PhysRevLett.100.180405} {\bibfield  {journal} {\bibinfo  {journal}
  {Phys. Rev. Lett.}\ }\textbf {\bibinfo {volume} {100}},\ \bibinfo {pages}
  {180405} (\bibinfo {year} {2008})}\BibitemShut {NoStop}%
\bibitem [{\citenamefont {Clad{\'e}}\ \emph {et~al.}(2009)\citenamefont
  {Clad{\'e}}, \citenamefont {Guellati-Kh{\'e}lifa}, \citenamefont {Nez},\ and\
  \citenamefont {Biraben}}]{Clade:2009}%
  \BibitemOpen
  \bibfield  {author} {\bibinfo {author} {\bibfnamefont {Pierre}\ \bibnamefont
  {Clad{\'e}}}, \bibinfo {author} {\bibfnamefont {Sa{\"\i}da}\ \bibnamefont
  {Guellati-Kh{\'e}lifa}}, \bibinfo {author} {\bibfnamefont {Fran{\c{c}}ois}\
  \bibnamefont {Nez}}, \ and\ \bibinfo {author} {\bibfnamefont
  {Fran{\c{c}}ois}\ \bibnamefont {Biraben}},\ }\bibfield  {title} {\enquote
  {\bibinfo {title} {Large momentum beam splitter using bloch oscillations},}\
  }\href@noop {} {\bibfield  {journal} {\bibinfo  {journal} {Phys. Rev. Lett.}\
  }\textbf {\bibinfo {volume} {102}},\ \bibinfo {pages} {240402} (\bibinfo
  {year} {2009})}\BibitemShut {NoStop}%
\bibitem [{\citenamefont {Chiow}\ \emph {et~al.}(2011)\citenamefont {Chiow},
  \citenamefont {Kovachy}, \citenamefont {Chien},\ and\ \citenamefont
  {Kasevich}}]{Chiow:2011}%
  \BibitemOpen
  \bibfield  {author} {\bibinfo {author} {\bibfnamefont {Sheng-wey}\
  \bibnamefont {Chiow}}, \bibinfo {author} {\bibfnamefont {Tim}\ \bibnamefont
  {Kovachy}}, \bibinfo {author} {\bibfnamefont {Hui-Chun}\ \bibnamefont
  {Chien}}, \ and\ \bibinfo {author} {\bibfnamefont {Mark~A.}\ \bibnamefont
  {Kasevich}},\ }\bibfield  {title} {\enquote {\bibinfo {title} {$102 \hbar k$
  large area atom interferometers},}\ }\href {\doibase
  10.1103/PhysRevLett.107.130403} {\bibfield  {journal} {\bibinfo  {journal}
  {Phys. Rev. Lett.}\ }\textbf {\bibinfo {volume} {107}},\ \bibinfo {pages}
  {130403} (\bibinfo {year} {2011})}\BibitemShut {NoStop}%
\bibitem [{\citenamefont {McDonald}\ \emph {et~al.}(2013)\citenamefont
  {McDonald}, \citenamefont {Kuhn}, \citenamefont {Bennetts}, \citenamefont
  {Debs}, \citenamefont {Hardman}, \citenamefont {Johnsson}, \citenamefont
  {Close},\ and\ \citenamefont {Robins}}]{McDonald:2013b}%
  \BibitemOpen
  \bibfield  {author} {\bibinfo {author} {\bibfnamefont {G.~D.}\ \bibnamefont
  {McDonald}}, \bibinfo {author} {\bibfnamefont {C.~C.~N.}\ \bibnamefont
  {Kuhn}}, \bibinfo {author} {\bibfnamefont {S.}~\bibnamefont {Bennetts}},
  \bibinfo {author} {\bibfnamefont {J.~E.}\ \bibnamefont {Debs}}, \bibinfo
  {author} {\bibfnamefont {K.~S.}\ \bibnamefont {Hardman}}, \bibinfo {author}
  {\bibfnamefont {M.}~\bibnamefont {Johnsson}}, \bibinfo {author}
  {\bibfnamefont {J.~D.}\ \bibnamefont {Close}}, \ and\ \bibinfo {author}
  {\bibfnamefont {N.~P.}\ \bibnamefont {Robins}},\ }\bibfield  {title}
  {\enquote {\bibinfo {title} {$80 \hbar k$ momentum separation with bloch
  oscillations in an optically guided atom interferometer},}\ }\href {\doibase
  10.1103/PhysRevA.88.053620} {\bibfield  {journal} {\bibinfo  {journal} {Phys.
  Rev. A}\ }\textbf {\bibinfo {volume} {88}},\ \bibinfo {pages} {053620}
  (\bibinfo {year} {2013})}\BibitemShut {NoStop}%
\bibitem [{\citenamefont {Mazzoni}\ \emph {et~al.}(2015)\citenamefont
  {Mazzoni}, \citenamefont {Zhang}, \citenamefont {Del~Aguila}, \citenamefont
  {Salvi}, \citenamefont {Poli},\ and\ \citenamefont {Tino}}]{Mazzoni:2015}%
  \BibitemOpen
  \bibfield  {author} {\bibinfo {author} {\bibfnamefont {T.}~\bibnamefont
  {Mazzoni}}, \bibinfo {author} {\bibfnamefont {X.}~\bibnamefont {Zhang}},
  \bibinfo {author} {\bibfnamefont {R.}~\bibnamefont {Del~Aguila}}, \bibinfo
  {author} {\bibfnamefont {L.}~\bibnamefont {Salvi}}, \bibinfo {author}
  {\bibfnamefont {N.}~\bibnamefont {Poli}}, \ and\ \bibinfo {author}
  {\bibfnamefont {G.~M.}\ \bibnamefont {Tino}},\ }\bibfield  {title} {\enquote
  {\bibinfo {title} {Large-momentum-transfer bragg interferometer with
  strontium atoms},}\ }\href {\doibase 10.1103/PhysRevA.92.053619} {\bibfield
  {journal} {\bibinfo  {journal} {Phys. Rev. A}\ }\textbf {\bibinfo {volume}
  {92}},\ \bibinfo {pages} {053619} (\bibinfo {year} {2015})}\BibitemShut
  {NoStop}%
\bibitem [{\citenamefont {Gross}\ \emph {et~al.}(2010)\citenamefont {Gross},
  \citenamefont {Zibold}, \citenamefont {Nicklas}, \citenamefont {Est{\`e}ve},\
  and\ \citenamefont {Oberthaler}}]{Gross:2010}%
  \BibitemOpen
  \bibfield  {author} {\bibinfo {author} {\bibfnamefont {C.}~\bibnamefont
  {Gross}}, \bibinfo {author} {\bibfnamefont {T.}~\bibnamefont {Zibold}},
  \bibinfo {author} {\bibfnamefont {E.}~\bibnamefont {Nicklas}}, \bibinfo
  {author} {\bibfnamefont {J.}~\bibnamefont {Est{\`e}ve}}, \ and\ \bibinfo
  {author} {\bibfnamefont {M.~K.}\ \bibnamefont {Oberthaler}},\ }\bibfield
  {title} {\enquote {\bibinfo {title} {Nonlinear atom interferometer surpasses
  classical precision limit},}\ }\href {http://dx.doi.org/10.1038/nature08919}
  {\bibfield  {journal} {\bibinfo  {journal} {Nature}\ }\textbf {\bibinfo
  {volume} {464}},\ \bibinfo {pages} {1165--1169} (\bibinfo {year}
  {2010})}\BibitemShut {NoStop}%
\bibitem [{\citenamefont {L{\"u}cke}\ \emph {et~al.}(2011)\citenamefont
  {L{\"u}cke}, \citenamefont {Scherer}, \citenamefont {Kruse}, \citenamefont
  {Pezz{\'e}}, \citenamefont {Deuretzbacher}, \citenamefont {Hyllus},
  \citenamefont {Topic}, \citenamefont {Peise}, \citenamefont {Ertmer},
  \citenamefont {Arlt}, \citenamefont {Santos}, \citenamefont {Smerzi},\ and\
  \citenamefont {Klempt}}]{Lucke:2011}%
  \BibitemOpen
  \bibfield  {author} {\bibinfo {author} {\bibfnamefont {B.}~\bibnamefont
  {L{\"u}cke}}, \bibinfo {author} {\bibfnamefont {M.}~\bibnamefont {Scherer}},
  \bibinfo {author} {\bibfnamefont {J.}~\bibnamefont {Kruse}}, \bibinfo
  {author} {\bibfnamefont {L.}~\bibnamefont {Pezz{\'e}}}, \bibinfo {author}
  {\bibfnamefont {F.}~\bibnamefont {Deuretzbacher}}, \bibinfo {author}
  {\bibfnamefont {P.}~\bibnamefont {Hyllus}}, \bibinfo {author} {\bibfnamefont
  {O.}~\bibnamefont {Topic}}, \bibinfo {author} {\bibfnamefont
  {J.}~\bibnamefont {Peise}}, \bibinfo {author} {\bibfnamefont
  {W.}~\bibnamefont {Ertmer}}, \bibinfo {author} {\bibfnamefont
  {J.}~\bibnamefont {Arlt}}, \bibinfo {author} {\bibfnamefont {L.}~\bibnamefont
  {Santos}}, \bibinfo {author} {\bibfnamefont {A.}~\bibnamefont {Smerzi}}, \
  and\ \bibinfo {author} {\bibfnamefont {C.}~\bibnamefont {Klempt}},\
  }\bibfield  {title} {\enquote {\bibinfo {title} {Twin matter waves for
  interferometry beyond the classical limit},}\ }\href@noop {} {\bibfield
  {journal} {\bibinfo  {journal} {Science}\ }\textbf {\bibinfo {volume}
  {334}},\ \bibinfo {pages} {773--776} (\bibinfo {year} {2011})}\BibitemShut
  {NoStop}%
\bibitem [{\citenamefont {Linnemann}\ \emph {et~al.}(2016)\citenamefont
  {Linnemann}, \citenamefont {Strobel}, \citenamefont {Muessel}, \citenamefont
  {Schulz}, \citenamefont {Lewis-Swan}, \citenamefont {Kheruntsyan},\ and\
  \citenamefont {Oberthaler}}]{Linnemann:2016}%
  \BibitemOpen
  \bibfield  {author} {\bibinfo {author} {\bibfnamefont {D.}~\bibnamefont
  {Linnemann}}, \bibinfo {author} {\bibfnamefont {H.}~\bibnamefont {Strobel}},
  \bibinfo {author} {\bibfnamefont {W.}~\bibnamefont {Muessel}}, \bibinfo
  {author} {\bibfnamefont {J.}~\bibnamefont {Schulz}}, \bibinfo {author}
  {\bibfnamefont {R.~J.}\ \bibnamefont {Lewis-Swan}}, \bibinfo {author}
  {\bibfnamefont {K.~V.}\ \bibnamefont {Kheruntsyan}}, \ and\ \bibinfo {author}
  {\bibfnamefont {M.~K.}\ \bibnamefont {Oberthaler}},\ }\bibfield  {title}
  {\enquote {\bibinfo {title} {Quantum-enhanced sensing based on time reversal
  of nonlinear dynamics},}\ }\href {\doibase 10.1103/PhysRevLett.117.013001}
  {\bibfield  {journal} {\bibinfo  {journal} {Phys. Rev. Lett.}\ }\textbf
  {\bibinfo {volume} {117}},\ \bibinfo {pages} {013001} (\bibinfo {year}
  {2016})}\BibitemShut {NoStop}%
\bibitem [{\citenamefont {Hosten}\ \emph {et~al.}(2016)\citenamefont {Hosten},
  \citenamefont {Engelsen}, \citenamefont {Krishnakumar},\ and\ \citenamefont
  {Kasevich}}]{Hosten:2016}%
  \BibitemOpen
  \bibfield  {author} {\bibinfo {author} {\bibfnamefont {Onur}\ \bibnamefont
  {Hosten}}, \bibinfo {author} {\bibfnamefont {Nils~J.}\ \bibnamefont
  {Engelsen}}, \bibinfo {author} {\bibfnamefont {Rajiv}\ \bibnamefont
  {Krishnakumar}}, \ and\ \bibinfo {author} {\bibfnamefont {Mark~A.}\
  \bibnamefont {Kasevich}},\ }\bibfield  {title} {\enquote {\bibinfo {title}
  {Measurement noise 100 times lower than the quantum-projection limit using
  entangled atoms},}\ }\href {http://dx.doi.org/10.1038/nature16176} {\bibfield
   {journal} {\bibinfo  {journal} {Nature}\ }\textbf {\bibinfo {volume}
  {529}},\ \bibinfo {pages} {505--508} (\bibinfo {year} {2016})}\BibitemShut
  {NoStop}%
\bibitem [{\citenamefont {Colangelo}\ \emph {et~al.}(2017)\citenamefont
  {Colangelo}, \citenamefont {Martin~Ciurana}, \citenamefont {Puentes},
  \citenamefont {Mitchell},\ and\ \citenamefont {Sewell}}]{Colangelo:2017}%
  \BibitemOpen
  \bibfield  {author} {\bibinfo {author} {\bibfnamefont {G.}~\bibnamefont
  {Colangelo}}, \bibinfo {author} {\bibfnamefont {F.}~\bibnamefont
  {Martin~Ciurana}}, \bibinfo {author} {\bibfnamefont {G.}~\bibnamefont
  {Puentes}}, \bibinfo {author} {\bibfnamefont {M.~W.}\ \bibnamefont
  {Mitchell}}, \ and\ \bibinfo {author} {\bibfnamefont {R.~J.}\ \bibnamefont
  {Sewell}},\ }\bibfield  {title} {\enquote {\bibinfo {title}
  {Entanglement-enhanced phase estimation without prior phase information},}\
  }\href {\doibase 10.1103/PhysRevLett.118.233603} {\bibfield  {journal}
  {\bibinfo  {journal} {Phys. Rev. Lett.}\ }\textbf {\bibinfo {volume} {118}},\
  \bibinfo {pages} {233603} (\bibinfo {year} {2017})}\BibitemShut {NoStop}%
\bibitem [{\citenamefont {Debs}\ \emph {et~al.}(2011)\citenamefont {Debs},
  \citenamefont {Altin}, \citenamefont {Barter}, \citenamefont {D\"oring},
  \citenamefont {Dennis}, \citenamefont {McDonald}, \citenamefont {Anderson},
  \citenamefont {Close},\ and\ \citenamefont {Robins}}]{Debs:2011}%
  \BibitemOpen
  \bibfield  {author} {\bibinfo {author} {\bibfnamefont {J.~E.}\ \bibnamefont
  {Debs}}, \bibinfo {author} {\bibfnamefont {P.~A.}\ \bibnamefont {Altin}},
  \bibinfo {author} {\bibfnamefont {T.~H.}\ \bibnamefont {Barter}}, \bibinfo
  {author} {\bibfnamefont {D.}~\bibnamefont {D\"oring}}, \bibinfo {author}
  {\bibfnamefont {G.~R.}\ \bibnamefont {Dennis}}, \bibinfo {author}
  {\bibfnamefont {G.}~\bibnamefont {McDonald}}, \bibinfo {author}
  {\bibfnamefont {R.~P.}\ \bibnamefont {Anderson}}, \bibinfo {author}
  {\bibfnamefont {J.~D.}\ \bibnamefont {Close}}, \ and\ \bibinfo {author}
  {\bibfnamefont {N.~P.}\ \bibnamefont {Robins}},\ }\bibfield  {title}
  {\enquote {\bibinfo {title} {Cold-atom gravimetry with a {B}ose-{E}instein
  condensate},}\ }\href {\doibase 10.1103/PhysRevA.84.033610} {\bibfield
  {journal} {\bibinfo  {journal} {Phys. Rev. A}\ }\textbf {\bibinfo {volume}
  {84}},\ \bibinfo {pages} {033610} (\bibinfo {year} {2011})}\BibitemShut
  {NoStop}%
\bibitem [{\citenamefont {Louchet-Chauvet}\ \emph {et~al.}(2011)\citenamefont
  {Louchet-Chauvet}, \citenamefont {Farah}, \citenamefont {Bodart},
  \citenamefont {Clairon}, \citenamefont {Landragin}, \citenamefont {Merlet},\
  and\ \citenamefont {{Pereira Dos Santos}}}]{Louchet-Chauvet:2011}%
  \BibitemOpen
  \bibfield  {author} {\bibinfo {author} {\bibfnamefont {Anne}\ \bibnamefont
  {Louchet-Chauvet}}, \bibinfo {author} {\bibfnamefont {Tristan}\ \bibnamefont
  {Farah}}, \bibinfo {author} {\bibfnamefont {Quentin}\ \bibnamefont {Bodart}},
  \bibinfo {author} {\bibfnamefont {Andr{\'e}}\ \bibnamefont {Clairon}},
  \bibinfo {author} {\bibfnamefont {Arnaud}\ \bibnamefont {Landragin}},
  \bibinfo {author} {\bibfnamefont {S{\'e}bastien}\ \bibnamefont {Merlet}}, \
  and\ \bibinfo {author} {\bibfnamefont {Franck}\ \bibnamefont {{Pereira Dos
  Santos}}},\ }\bibfield  {title} {\enquote {\bibinfo {title} {The influence of
  transverse motion within an atomic gravimeter},}\ }\href
  {http://stacks.iop.org/1367-2630/13/i=6/a=065025} {\bibfield  {journal}
  {\bibinfo  {journal} {New Journal of Physics}\ }\textbf {\bibinfo {volume}
  {13}},\ \bibinfo {pages} {065025} (\bibinfo {year} {2011})}\BibitemShut
  {NoStop}%
\bibitem [{\citenamefont {Szigeti}\ \emph {et~al.}(2012)\citenamefont
  {Szigeti}, \citenamefont {Debs}, \citenamefont {Hope}, \citenamefont
  {Robins},\ and\ \citenamefont {Close}}]{Szigeti:2012}%
  \BibitemOpen
  \bibfield  {author} {\bibinfo {author} {\bibfnamefont {S~S}\ \bibnamefont
  {Szigeti}}, \bibinfo {author} {\bibfnamefont {J~E}\ \bibnamefont {Debs}},
  \bibinfo {author} {\bibfnamefont {J~J}\ \bibnamefont {Hope}}, \bibinfo
  {author} {\bibfnamefont {N~P}\ \bibnamefont {Robins}}, \ and\ \bibinfo
  {author} {\bibfnamefont {J~D}\ \bibnamefont {Close}},\ }\bibfield  {title}
  {\enquote {\bibinfo {title} {Why momentum width matters for atom
  interferometry with bragg pulses},}\ }\href
  {http://stacks.iop.org/1367-2630/14/i=2/a=023009} {\bibfield  {journal}
  {\bibinfo  {journal} {New Journal of Physics}\ }\textbf {\bibinfo {volume}
  {14}},\ \bibinfo {pages} {023009} (\bibinfo {year} {2012})}\BibitemShut
  {NoStop}%
\bibitem [{\citenamefont {Hardman}\ \emph {et~al.}(2014)\citenamefont
  {Hardman}, \citenamefont {Kuhn}, \citenamefont {McDonald}, \citenamefont
  {Debs}, \citenamefont {Bennetts}, \citenamefont {Close},\ and\ \citenamefont
  {Robins}}]{Hardman:2014}%
  \BibitemOpen
  \bibfield  {author} {\bibinfo {author} {\bibfnamefont {Kyle~S.}\ \bibnamefont
  {Hardman}}, \bibinfo {author} {\bibfnamefont {Carlos C.~N.}\ \bibnamefont
  {Kuhn}}, \bibinfo {author} {\bibfnamefont {Gordon~D.}\ \bibnamefont
  {McDonald}}, \bibinfo {author} {\bibfnamefont {John~E.}\ \bibnamefont
  {Debs}}, \bibinfo {author} {\bibfnamefont {Shayne}\ \bibnamefont {Bennetts}},
  \bibinfo {author} {\bibfnamefont {John~D.}\ \bibnamefont {Close}}, \ and\
  \bibinfo {author} {\bibfnamefont {Nicholas~P.}\ \bibnamefont {Robins}},\
  }\bibfield  {title} {\enquote {\bibinfo {title} {Role of source coherence in
  atom interferometery},}\ }\href {\doibase 10.1103/PhysRevA.89.023626}
  {\bibfield  {journal} {\bibinfo  {journal} {Phys. Rev. A}\ }\textbf {\bibinfo
  {volume} {89}},\ \bibinfo {pages} {023626} (\bibinfo {year}
  {2014})}\BibitemShut {NoStop}%
\bibitem [{\citenamefont {Robins}\ \emph {et~al.}(2013)\citenamefont {Robins},
  \citenamefont {Altin}, \citenamefont {Debs},\ and\ \citenamefont
  {Close}}]{Robins:2013}%
  \BibitemOpen
  \bibfield  {author} {\bibinfo {author} {\bibfnamefont {N.~P.}\ \bibnamefont
  {Robins}}, \bibinfo {author} {\bibfnamefont {P.~A.}\ \bibnamefont {Altin}},
  \bibinfo {author} {\bibfnamefont {J.~E.}\ \bibnamefont {Debs}}, \ and\
  \bibinfo {author} {\bibfnamefont {J.~D.}\ \bibnamefont {Close}},\ }\bibfield
  {title} {\enquote {\bibinfo {title} {Atom lasers: Production, properties and
  prospects for precision inertial measurement},}\ }\href {\doibase
  http://dx.doi.org/10.1016/j.physrep.2013.03.006} {\bibfield  {journal}
  {\bibinfo  {journal} {Physics Reports}\ }\textbf {\bibinfo {volume} {529}},\
  \bibinfo {pages} {265--296} (\bibinfo {year} {2013})}\BibitemShut {NoStop}%
\bibitem [{\citenamefont {Szigeti}\ \emph {et~al.}(2009)\citenamefont
  {Szigeti}, \citenamefont {Hush}, \citenamefont {Carvalho},\ and\
  \citenamefont {Hope}}]{Szigeti:2009}%
  \BibitemOpen
  \bibfield  {author} {\bibinfo {author} {\bibfnamefont {S.~S.}\ \bibnamefont
  {Szigeti}}, \bibinfo {author} {\bibfnamefont {M.~R.}\ \bibnamefont {Hush}},
  \bibinfo {author} {\bibfnamefont {A.~R.~R.}\ \bibnamefont {Carvalho}}, \ and\
  \bibinfo {author} {\bibfnamefont {J.~J.}\ \bibnamefont {Hope}},\ }\bibfield
  {title} {\enquote {\bibinfo {title} {Continuous measurement feedback control
  of a bose-einstein condensate using phase-contrast imaging},}\ }\href
  {\doibase 10.1103/PhysRevA.80.013614} {\bibfield  {journal} {\bibinfo
  {journal} {Physical Review A}\ }\textbf {\bibinfo {volume} {80}},\ \bibinfo
  {eid} {013614} (\bibinfo {year} {2009})}\BibitemShut {NoStop}%
\bibitem [{\citenamefont {Szigeti}\ \emph {et~al.}(2010)\citenamefont
  {Szigeti}, \citenamefont {Hush}, \citenamefont {Carvalho},\ and\
  \citenamefont {Hope}}]{Szigeti:2010}%
  \BibitemOpen
  \bibfield  {author} {\bibinfo {author} {\bibfnamefont {S.~S.}\ \bibnamefont
  {Szigeti}}, \bibinfo {author} {\bibfnamefont {M.~R.}\ \bibnamefont {Hush}},
  \bibinfo {author} {\bibfnamefont {A.~R.~R.}\ \bibnamefont {Carvalho}}, \ and\
  \bibinfo {author} {\bibfnamefont {J.~J.}\ \bibnamefont {Hope}},\ }\bibfield
  {title} {\enquote {\bibinfo {title} {Feedback control of an interacting
  bose-einstein condensate using phase-contrast imaging},}\ }\href {\doibase
  10.1103/PhysRevA.82.043632} {\bibfield  {journal} {\bibinfo  {journal} {Phys.
  Rev. A}\ }\textbf {\bibinfo {volume} {82}},\ \bibinfo {pages} {043632}
  (\bibinfo {year} {2010})}\BibitemShut {NoStop}%
\bibitem [{\citenamefont {Hush}\ \emph {et~al.}(2013)\citenamefont {Hush},
  \citenamefont {Szigeti}, \citenamefont {Carvalho},\ and\ \citenamefont
  {Hope}}]{Hush:2013}%
  \BibitemOpen
  \bibfield  {author} {\bibinfo {author} {\bibfnamefont {M~R}\ \bibnamefont
  {Hush}}, \bibinfo {author} {\bibfnamefont {S~S}\ \bibnamefont {Szigeti}},
  \bibinfo {author} {\bibfnamefont {A~R~R}\ \bibnamefont {Carvalho}}, \ and\
  \bibinfo {author} {\bibfnamefont {J~J}\ \bibnamefont {Hope}},\ }\bibfield
  {title} {\enquote {\bibinfo {title} {Controlling spontaneous-emission noise
  in measurement-based feedback cooling of a bose--einstein condensate},}\
  }\href {http://stacks.iop.org/1367-2630/15/i=11/a=113060} {\bibfield
  {journal} {\bibinfo  {journal} {New Journal of Physics}\ }\textbf {\bibinfo
  {volume} {15}},\ \bibinfo {pages} {113060} (\bibinfo {year}
  {2013})}\BibitemShut {NoStop}%
\bibitem [{\citenamefont {Altin}\ \emph {et~al.}(2011)\citenamefont {Altin},
  \citenamefont {McDonald}, \citenamefont {Doring}, \citenamefont {Debs},
  \citenamefont {Barter}, \citenamefont {Close}, \citenamefont {Robins},
  \citenamefont {Haine}, \citenamefont {Hanna},\ and\ \citenamefont
  {Anderson}}]{Altin:2011}%
  \BibitemOpen
  \bibfield  {author} {\bibinfo {author} {\bibfnamefont {P~A}\ \bibnamefont
  {Altin}}, \bibinfo {author} {\bibfnamefont {G}~\bibnamefont {McDonald}},
  \bibinfo {author} {\bibfnamefont {D}~\bibnamefont {Doring}}, \bibinfo
  {author} {\bibfnamefont {J~E}\ \bibnamefont {Debs}}, \bibinfo {author}
  {\bibfnamefont {T~H}\ \bibnamefont {Barter}}, \bibinfo {author}
  {\bibfnamefont {J~D}\ \bibnamefont {Close}}, \bibinfo {author} {\bibfnamefont
  {N~P}\ \bibnamefont {Robins}}, \bibinfo {author} {\bibfnamefont {S~A}\
  \bibnamefont {Haine}}, \bibinfo {author} {\bibfnamefont {T~M}\ \bibnamefont
  {Hanna}}, \ and\ \bibinfo {author} {\bibfnamefont {R~P}\ \bibnamefont
  {Anderson}},\ }\bibfield  {title} {\enquote {\bibinfo {title} {Optically
  trapped atom interferometry using the clock transition of large 87 rb
  bose-einstein condensates},}\ }\href
  {http://stacks.iop.org/1367-2630/13/i=6/a=065020} {\bibfield  {journal}
  {\bibinfo  {journal} {New Journal of Physics}\ }\textbf {\bibinfo {volume}
  {13}},\ \bibinfo {pages} {065020} (\bibinfo {year} {2011})}\BibitemShut
  {NoStop}%
\bibitem [{\citenamefont {Haine}\ and\ \citenamefont
  {Ferris}(2011)}]{Haine:2011}%
  \BibitemOpen
  \bibfield  {author} {\bibinfo {author} {\bibfnamefont {S.~A.}\ \bibnamefont
  {Haine}}\ and\ \bibinfo {author} {\bibfnamefont {A.~J.}\ \bibnamefont
  {Ferris}},\ }\bibfield  {title} {\enquote {\bibinfo {title} {Surpassing the
  standard quantum limit in an atom interferometer with four-mode entanglement
  produced from four-wave mixing},}\ }\href {\doibase
  10.1103/PhysRevA.84.043624} {\bibfield  {journal} {\bibinfo  {journal} {Phys.
  Rev. A}\ }\textbf {\bibinfo {volume} {84}},\ \bibinfo {pages} {043624}
  (\bibinfo {year} {2011})}\BibitemShut {NoStop}%
\bibitem [{\citenamefont {Haine}(2013)}]{Haine:2013}%
  \BibitemOpen
  \bibfield  {author} {\bibinfo {author} {\bibfnamefont {S.~A.}\ \bibnamefont
  {Haine}},\ }\bibfield  {title} {\enquote {\bibinfo {title}
  {Information-recycling beam splitters for quantum enhanced atom
  interferometry},}\ }\href {\doibase 10.1103/PhysRevLett.110.053002}
  {\bibfield  {journal} {\bibinfo  {journal} {Phys. Rev. Lett.}\ }\textbf
  {\bibinfo {volume} {110}},\ \bibinfo {pages} {053002} (\bibinfo {year}
  {2013})}\BibitemShut {NoStop}%
\bibitem [{\citenamefont {Haine}\ \emph {et~al.}(2014)\citenamefont {Haine},
  \citenamefont {Lau}, \citenamefont {Anderson},\ and\ \citenamefont
  {Johnsson}}]{Haine:2014}%
  \BibitemOpen
  \bibfield  {author} {\bibinfo {author} {\bibfnamefont {S.~A.}\ \bibnamefont
  {Haine}}, \bibinfo {author} {\bibfnamefont {J.}~\bibnamefont {Lau}}, \bibinfo
  {author} {\bibfnamefont {R.~P.}\ \bibnamefont {Anderson}}, \ and\ \bibinfo
  {author} {\bibfnamefont {M.~T.}\ \bibnamefont {Johnsson}},\ }\bibfield
  {title} {\enquote {\bibinfo {title} {Self-induced spatial dynamics to enhance
  spin squeezing via one-axis twisting in a two-component bose-einstein
  condensate},}\ }\href {\doibase 10.1103/PhysRevA.90.023613} {\bibfield
  {journal} {\bibinfo  {journal} {Phys. Rev. A}\ }\textbf {\bibinfo {volume}
  {90}},\ \bibinfo {pages} {023613} (\bibinfo {year} {2014})}\BibitemShut
  {NoStop}%
\bibitem [{\citenamefont {Szigeti}\ \emph {et~al.}(2014)\citenamefont
  {Szigeti}, \citenamefont {Tonekaboni}, \citenamefont {Lau}, \citenamefont
  {Hood},\ and\ \citenamefont {Haine}}]{Szigeti:2014b}%
  \BibitemOpen
  \bibfield  {author} {\bibinfo {author} {\bibfnamefont {Stuart~S.}\
  \bibnamefont {Szigeti}}, \bibinfo {author} {\bibfnamefont {Behnam}\
  \bibnamefont {Tonekaboni}}, \bibinfo {author} {\bibfnamefont {Wing Yung~S.}\
  \bibnamefont {Lau}}, \bibinfo {author} {\bibfnamefont {Samantha~N.}\
  \bibnamefont {Hood}}, \ and\ \bibinfo {author} {\bibfnamefont {Simon~A.}\
  \bibnamefont {Haine}},\ }\bibfield  {title} {\enquote {\bibinfo {title}
  {Squeezed-light-enhanced atom interferometry below the standard quantum
  limit},}\ }\href {\doibase 10.1103/PhysRevA.90.063630} {\bibfield  {journal}
  {\bibinfo  {journal} {Phys. Rev. A}\ }\textbf {\bibinfo {volume} {90}},\
  \bibinfo {pages} {063630} (\bibinfo {year} {2014})}\BibitemShut {NoStop}%
\bibitem [{\citenamefont {Tonekaboni}\ \emph {et~al.}(2015)\citenamefont
  {Tonekaboni}, \citenamefont {Haine},\ and\ \citenamefont
  {Szigeti}}]{Tonekaboni:2015}%
  \BibitemOpen
  \bibfield  {author} {\bibinfo {author} {\bibfnamefont {Behnam}\ \bibnamefont
  {Tonekaboni}}, \bibinfo {author} {\bibfnamefont {Simon~A.}\ \bibnamefont
  {Haine}}, \ and\ \bibinfo {author} {\bibfnamefont {Stuart~S.}\ \bibnamefont
  {Szigeti}},\ }\bibfield  {title} {\enquote {\bibinfo {title}
  {Heisenberg-limited metrology with a squeezed vacuum state, three-mode
  mixing, and information recycling},}\ }\href {\doibase
  10.1103/PhysRevA.91.033616} {\bibfield  {journal} {\bibinfo  {journal} {Phys.
  Rev. A}\ }\textbf {\bibinfo {volume} {91}},\ \bibinfo {pages} {033616}
  (\bibinfo {year} {2015})}\BibitemShut {NoStop}%
\bibitem [{\citenamefont {Haine}\ \emph {et~al.}(2015)\citenamefont {Haine},
  \citenamefont {Szigeti}, \citenamefont {Lang},\ and\ \citenamefont
  {Caves}}]{Haine:2015}%
  \BibitemOpen
  \bibfield  {author} {\bibinfo {author} {\bibfnamefont {Simon~A.}\
  \bibnamefont {Haine}}, \bibinfo {author} {\bibfnamefont {Stuart~S.}\
  \bibnamefont {Szigeti}}, \bibinfo {author} {\bibfnamefont {Matthias~D.}\
  \bibnamefont {Lang}}, \ and\ \bibinfo {author} {\bibfnamefont {Carlton~M.}\
  \bibnamefont {Caves}},\ }\bibfield  {title} {\enquote {\bibinfo {title}
  {Heisenberg-limited metrology with information recycling},}\ }\href {\doibase
  10.1103/PhysRevA.91.041802} {\bibfield  {journal} {\bibinfo  {journal} {Phys.
  Rev. A}\ }\textbf {\bibinfo {volume} {91}},\ \bibinfo {pages} {041802}
  (\bibinfo {year} {2015})}\BibitemShut {NoStop}%
\bibitem [{\citenamefont {Haine}\ and\ \citenamefont
  {Szigeti}(2015)}]{Haine:2015b}%
  \BibitemOpen
  \bibfield  {author} {\bibinfo {author} {\bibfnamefont {Simon~A.}\
  \bibnamefont {Haine}}\ and\ \bibinfo {author} {\bibfnamefont {Stuart~S.}\
  \bibnamefont {Szigeti}},\ }\bibfield  {title} {\enquote {\bibinfo {title}
  {Quantum metrology with mixed states: When recovering lost information is
  better than never losing it},}\ }\href {\doibase 10.1103/PhysRevA.92.032317}
  {\bibfield  {journal} {\bibinfo  {journal} {Phys. Rev. A}\ }\textbf {\bibinfo
  {volume} {92}},\ \bibinfo {pages} {032317} (\bibinfo {year}
  {2015})}\BibitemShut {NoStop}%
\bibitem [{\citenamefont {Nolan}\ \emph {et~al.}(2016)\citenamefont {Nolan},
  \citenamefont {Sabbatini}, \citenamefont {Bromley}, \citenamefont {Davis},\
  and\ \citenamefont {Haine}}]{Nolan:2016}%
  \BibitemOpen
  \bibfield  {author} {\bibinfo {author} {\bibfnamefont {Samuel~P.}\
  \bibnamefont {Nolan}}, \bibinfo {author} {\bibfnamefont {Jacopo}\
  \bibnamefont {Sabbatini}}, \bibinfo {author} {\bibfnamefont {Michael W.~J.}\
  \bibnamefont {Bromley}}, \bibinfo {author} {\bibfnamefont {Matthew~J.}\
  \bibnamefont {Davis}}, \ and\ \bibinfo {author} {\bibfnamefont {Simon~A.}\
  \bibnamefont {Haine}},\ }\bibfield  {title} {\enquote {\bibinfo {title}
  {Quantum enhanced measurement of rotations with a spin-1 bose-einstein
  condensate in a ring trap},}\ }\href {\doibase 10.1103/PhysRevA.93.023616}
  {\bibfield  {journal} {\bibinfo  {journal} {Phys. Rev. A}\ }\textbf {\bibinfo
  {volume} {93}},\ \bibinfo {pages} {023616} (\bibinfo {year}
  {2016})}\BibitemShut {NoStop}%
\bibitem [{\citenamefont {Haine}\ and\ \citenamefont {Lau}(2016)}]{Haine:2016}%
  \BibitemOpen
  \bibfield  {author} {\bibinfo {author} {\bibfnamefont {Simon~A.}\
  \bibnamefont {Haine}}\ and\ \bibinfo {author} {\bibfnamefont {Wing
  Yung~Sarah}\ \bibnamefont {Lau}},\ }\bibfield  {title} {\enquote {\bibinfo
  {title} {Generation of atom-light entanglement in an optical cavity for
  quantum enhanced atom interferometry},}\ }\href {\doibase
  10.1103/PhysRevA.93.023607} {\bibfield  {journal} {\bibinfo  {journal} {Phys.
  Rev. A}\ }\textbf {\bibinfo {volume} {93}},\ \bibinfo {pages} {023607}
  (\bibinfo {year} {2016})}\BibitemShut {NoStop}%
\bibitem [{\citenamefont {Szigeti}\ \emph {et~al.}(2017)\citenamefont
  {Szigeti}, \citenamefont {Lewis-Swan},\ and\ \citenamefont
  {Haine}}]{Szigeti:2017}%
  \BibitemOpen
  \bibfield  {author} {\bibinfo {author} {\bibfnamefont {Stuart~S.}\
  \bibnamefont {Szigeti}}, \bibinfo {author} {\bibfnamefont {Robert~J.}\
  \bibnamefont {Lewis-Swan}}, \ and\ \bibinfo {author} {\bibfnamefont
  {Simon~A.}\ \bibnamefont {Haine}},\ }\bibfield  {title} {\enquote {\bibinfo
  {title} {Pumped-up su(1,1) interferometry},}\ }\href {\doibase
  10.1103/PhysRevLett.118.150401} {\bibfield  {journal} {\bibinfo  {journal}
  {Phys. Rev. Lett.}\ }\textbf {\bibinfo {volume} {118}},\ \bibinfo {pages}
  {150401} (\bibinfo {year} {2017})}\BibitemShut {NoStop}%
\bibitem [{\citenamefont {Haine}(2018{\natexlab{a}})}]{Haine:2018}%
  \BibitemOpen
  \bibfield  {author} {\bibinfo {author} {\bibfnamefont {Simon~A}\ \bibnamefont
  {Haine}},\ }\bibfield  {title} {\enquote {\bibinfo {title} {Quantum noise in
  bright soliton matterwave interferometry},}\ }\href
  {http://stacks.iop.org/1367-2630/20/i=3/a=033009} {\bibfield  {journal}
  {\bibinfo  {journal} {New Journal of Physics}\ }\textbf {\bibinfo {volume}
  {20}},\ \bibinfo {pages} {033009} (\bibinfo {year}
  {2018}{\natexlab{a}})}\BibitemShut {NoStop}%
\bibitem [{\citenamefont {Lan}\ \emph {et~al.}(2012)\citenamefont {Lan},
  \citenamefont {Kuan}, \citenamefont {Estey}, \citenamefont {Haslinger},\ and\
  \citenamefont {M\"uller}}]{Lan:2012}%
  \BibitemOpen
  \bibfield  {author} {\bibinfo {author} {\bibfnamefont {Shau-Yu}\ \bibnamefont
  {Lan}}, \bibinfo {author} {\bibfnamefont {Pei-Chen}\ \bibnamefont {Kuan}},
  \bibinfo {author} {\bibfnamefont {Brian}\ \bibnamefont {Estey}}, \bibinfo
  {author} {\bibfnamefont {Philipp}\ \bibnamefont {Haslinger}}, \ and\ \bibinfo
  {author} {\bibfnamefont {Holger}\ \bibnamefont {M\"uller}},\ }\bibfield
  {title} {\enquote {\bibinfo {title} {Influence of the coriolis force in atom
  interferometry},}\ }\href {\doibase 10.1103/PhysRevLett.108.090402}
  {\bibfield  {journal} {\bibinfo  {journal} {Phys. Rev. Lett.}\ }\textbf
  {\bibinfo {volume} {108}},\ \bibinfo {pages} {090402} (\bibinfo {year}
  {2012})}\BibitemShut {NoStop}%
\bibitem [{\citenamefont {Schkolnik}\ \emph {et~al.}(2015)\citenamefont
  {Schkolnik}, \citenamefont {Leykauf}, \citenamefont {Hauth}, \citenamefont
  {Freier},\ and\ \citenamefont {Peters}}]{Schkolnik:2015}%
  \BibitemOpen
  \bibfield  {author} {\bibinfo {author} {\bibfnamefont {V.}~\bibnamefont
  {Schkolnik}}, \bibinfo {author} {\bibfnamefont {B.}~\bibnamefont {Leykauf}},
  \bibinfo {author} {\bibfnamefont {M.}~\bibnamefont {Hauth}}, \bibinfo
  {author} {\bibfnamefont {C.}~\bibnamefont {Freier}}, \ and\ \bibinfo {author}
  {\bibfnamefont {A.}~\bibnamefont {Peters}},\ }\bibfield  {title} {\enquote
  {\bibinfo {title} {The effect of wavefront aberrations in atom
  interferometry},}\ }\href {\doibase 10.1007/s00340-015-6138-5} {\bibfield
  {journal} {\bibinfo  {journal} {Applied Physics B}\ }\textbf {\bibinfo
  {volume} {120}},\ \bibinfo {pages} {311--316} (\bibinfo {year}
  {2015})}\BibitemShut {NoStop}%
\bibitem [{\citenamefont {Braunstein}\ and\ \citenamefont
  {Caves}(1994)}]{Braunstein:1994}%
  \BibitemOpen
  \bibfield  {author} {\bibinfo {author} {\bibfnamefont {Samuel~L.}\
  \bibnamefont {Braunstein}}\ and\ \bibinfo {author} {\bibfnamefont
  {Carlton~M.}\ \bibnamefont {Caves}},\ }\bibfield  {title} {\enquote {\bibinfo
  {title} {Statistical distance and the geometry of quantum states},}\ }\href
  {\doibase 10.1103/PhysRevLett.72.3439} {\bibfield  {journal} {\bibinfo
  {journal} {Phys. Rev. Lett.}\ }\textbf {\bibinfo {volume} {72}},\ \bibinfo
  {pages} {3439--3443} (\bibinfo {year} {1994})}\BibitemShut {NoStop}%
\bibitem [{\citenamefont {Demkowicz-Dobrza{\'n}ski}\ \emph
  {et~al.}(2015)\citenamefont {Demkowicz-Dobrza{\'n}ski}, \citenamefont
  {Jarzyna},\ and\ \citenamefont
  {Ko{\l}ody{\'n}ski}}]{Demkowicz-Dobrzanski:2014}%
  \BibitemOpen
  \bibfield  {author} {\bibinfo {author} {\bibfnamefont {Rafa{\l}}\
  \bibnamefont {Demkowicz-Dobrza{\'n}ski}}, \bibinfo {author} {\bibfnamefont
  {M.}~\bibnamefont {Jarzyna}}, \ and\ \bibinfo {author} {\bibfnamefont
  {J.}~\bibnamefont {Ko{\l}ody{\'n}ski}},\ }\bibfield  {title} {\enquote
  {\bibinfo {title} {Quantum limits in optical interferometry},}\ }\href@noop
  {} {\bibfield  {journal} {\bibinfo  {journal} {Progress in Optics}\ }\textbf
  {\bibinfo {volume} {345}} (\bibinfo {year} {2015})}\BibitemShut {NoStop}%
\bibitem [{\citenamefont {T\'oth}\ and\ \citenamefont
  {Apellaniz}(2014)}]{Toth:2014}%
  \BibitemOpen
  \bibfield  {author} {\bibinfo {author} {\bibfnamefont {G\'eza}\ \bibnamefont
  {T\'oth}}\ and\ \bibinfo {author} {\bibfnamefont {Iagoba}\ \bibnamefont
  {Apellaniz}},\ }\bibfield  {title} {\enquote {\bibinfo {title} {Quantum
  metrology from a quantum information science perspective},}\ }\href
  {http://stacks.iop.org/1751-8121/47/i=42/a=424006} {\bibfield  {journal}
  {\bibinfo  {journal} {Journal of Physics A: Mathematical and Theoretical}\
  }\textbf {\bibinfo {volume} {47}},\ \bibinfo {pages} {424006} (\bibinfo
  {year} {2014})}\BibitemShut {NoStop}%
\bibitem [{\citenamefont {Haine}(2016)}]{Haine:2016b}%
  \BibitemOpen
  \bibfield  {author} {\bibinfo {author} {\bibfnamefont {Simon~A.}\
  \bibnamefont {Haine}},\ }\bibfield  {title} {\enquote {\bibinfo {title}
  {Mean-field dynamics and fisher information in matter wave interferometry},}\
  }\href {\doibase 10.1103/PhysRevLett.116.230404} {\bibfield  {journal}
  {\bibinfo  {journal} {Phys. Rev. Lett.}\ }\textbf {\bibinfo {volume} {116}},\
  \bibinfo {pages} {230404} (\bibinfo {year} {2016})}\BibitemShut {NoStop}%
\bibitem [{\citenamefont {Qvarfort}\ \emph {et~al.}((2017))\citenamefont
  {Qvarfort}, \citenamefont {Serafini}, \citenamefont {Barker},\ and\
  \citenamefont {Bose}}]{Qvarfort:2017}%
  \BibitemOpen
  \bibfield  {author} {\bibinfo {author} {\bibfnamefont {Sofia}\ \bibnamefont
  {Qvarfort}}, \bibinfo {author} {\bibfnamefont {Alessio}\ \bibnamefont
  {Serafini}}, \bibinfo {author} {\bibfnamefont {Peter}\ \bibnamefont
  {Barker}}, \ and\ \bibinfo {author} {\bibfnamefont {Sougato}\ \bibnamefont
  {Bose}},\ }\bibfield  {title} {\enquote {\bibinfo {title} {Gravimetry through
  non-linear optomechanics},}\ }\href@noop {} {\bibfield  {journal} {\bibinfo
  {journal} {arXiv:1706.09131}\ } (\bibinfo {year} {(2017)})}\BibitemShut
  {NoStop}%
\bibitem [{\citenamefont {Matthews}\ \emph {et~al.}(1999)\citenamefont
  {Matthews}, \citenamefont {Anderson}, \citenamefont {Haljan}, \citenamefont
  {Hall}, \citenamefont {Holland}, \citenamefont {Williams}, \citenamefont
  {Wieman},\ and\ \citenamefont {Cornell}}]{Matthews:1999}%
  \BibitemOpen
  \bibfield  {author} {\bibinfo {author} {\bibfnamefont {M.~R.}\ \bibnamefont
  {Matthews}}, \bibinfo {author} {\bibfnamefont {B.~P.}\ \bibnamefont
  {Anderson}}, \bibinfo {author} {\bibfnamefont {P.~C.}\ \bibnamefont
  {Haljan}}, \bibinfo {author} {\bibfnamefont {D.~S.}\ \bibnamefont {Hall}},
  \bibinfo {author} {\bibfnamefont {M.~J.}\ \bibnamefont {Holland}}, \bibinfo
  {author} {\bibfnamefont {J.~E.}\ \bibnamefont {Williams}}, \bibinfo {author}
  {\bibfnamefont {C.~E.}\ \bibnamefont {Wieman}}, \ and\ \bibinfo {author}
  {\bibfnamefont {E.~A.}\ \bibnamefont {Cornell}},\ }\bibfield  {title}
  {\enquote {\bibinfo {title} {Watching a superfluid untwist itself: Recurrence
  of rabi oscillations in a bose-einstein condensate},}\ }\href {\doibase
  10.1103/PhysRevLett.83.3358} {\bibfield  {journal} {\bibinfo  {journal}
  {Phys. Rev. Lett.}\ }\textbf {\bibinfo {volume} {83}},\ \bibinfo {pages}
  {3358--3361} (\bibinfo {year} {1999})}\BibitemShut {NoStop}%
\bibitem [{\citenamefont {Ammann}\ and\ \citenamefont
  {Christensen}(1997)}]{Ammann:1997}%
  \BibitemOpen
  \bibfield  {author} {\bibinfo {author} {\bibfnamefont {Hubert}\ \bibnamefont
  {Ammann}}\ and\ \bibinfo {author} {\bibfnamefont {Nelson}\ \bibnamefont
  {Christensen}},\ }\bibfield  {title} {\enquote {\bibinfo {title} {Delta kick
  cooling: A new method for cooling atoms},}\ }\href {\doibase
  10.1103/PhysRevLett.78.2088} {\bibfield  {journal} {\bibinfo  {journal}
  {Phys. Rev. Lett.}\ }\textbf {\bibinfo {volume} {78}},\ \bibinfo {pages}
  {2088--2091} (\bibinfo {year} {1997})}\BibitemShut {NoStop}%
\bibitem [{\citenamefont {Pezz\'e}\ and\ \citenamefont
  {Smerzi}(2013)}]{Pezze:2013}%
  \BibitemOpen
  \bibfield  {author} {\bibinfo {author} {\bibfnamefont {Luca}\ \bibnamefont
  {Pezz\'e}}\ and\ \bibinfo {author} {\bibfnamefont {Augusto}\ \bibnamefont
  {Smerzi}},\ }\bibfield  {title} {\enquote {\bibinfo {title} {Ultrasensitive
  two-mode interferometry with single-mode number squeezing},}\ }\href
  {\doibase 10.1103/PhysRevLett.110.163604} {\bibfield  {journal} {\bibinfo
  {journal} {Phys. Rev. Lett.}\ }\textbf {\bibinfo {volume} {110}},\ \bibinfo
  {pages} {163604} (\bibinfo {year} {2013})}\BibitemShut {NoStop}%
\bibitem [{\citenamefont {Gabbrielli}\ \emph {et~al.}(2015)\citenamefont
  {Gabbrielli}, \citenamefont {Pezz\`e},\ and\ \citenamefont
  {Smerzi}}]{Gabbrielli:2015}%
  \BibitemOpen
  \bibfield  {author} {\bibinfo {author} {\bibfnamefont {Marco}\ \bibnamefont
  {Gabbrielli}}, \bibinfo {author} {\bibfnamefont {Luca}\ \bibnamefont
  {Pezz\`e}}, \ and\ \bibinfo {author} {\bibfnamefont {Augusto}\ \bibnamefont
  {Smerzi}},\ }\bibfield  {title} {\enquote {\bibinfo {title} {Spin-mixing
  interferometry with bose-einstein condensates},}\ }\href {\doibase
  10.1103/PhysRevLett.115.163002} {\bibfield  {journal} {\bibinfo  {journal}
  {Phys. Rev. Lett.}\ }\textbf {\bibinfo {volume} {115}},\ \bibinfo {pages}
  {163002} (\bibinfo {year} {2015})}\BibitemShut {NoStop}%
\bibitem [{\citenamefont {Nolan}\ \emph {et~al.}(2017)\citenamefont {Nolan},
  \citenamefont {Szigeti},\ and\ \citenamefont {Haine}}]{Nolan:2017b}%
  \BibitemOpen
  \bibfield  {author} {\bibinfo {author} {\bibfnamefont {Samuel~P.}\
  \bibnamefont {Nolan}}, \bibinfo {author} {\bibfnamefont {Stuart~S.}\
  \bibnamefont {Szigeti}}, \ and\ \bibinfo {author} {\bibfnamefont {Simon~A.}\
  \bibnamefont {Haine}},\ }\bibfield  {title} {\enquote {\bibinfo {title}
  {Optimal and robust quantum metrology using interaction-based readouts},}\
  }\href {\doibase 10.1103/PhysRevLett.119.193601} {\bibfield  {journal}
  {\bibinfo  {journal} {Phys. Rev. Lett.}\ }\textbf {\bibinfo {volume} {119}},\
  \bibinfo {pages} {193601} (\bibinfo {year} {2017})}\BibitemShut {NoStop}%
\bibitem [{\citenamefont {Mirkhalaf}\ \emph {et~al.}(2018)\citenamefont
  {Mirkhalaf}, \citenamefont {Nolan},\ and\ \citenamefont
  {Haine}}]{Mirkhalaf:2018}%
  \BibitemOpen
  \bibfield  {author} {\bibinfo {author} {\bibfnamefont {S.~S.}\ \bibnamefont
  {Mirkhalaf}}, \bibinfo {author} {\bibfnamefont {S.~P.}\ \bibnamefont
  {Nolan}}, \ and\ \bibinfo {author} {\bibfnamefont {S.~A.}\ \bibnamefont
  {Haine}},\ }\bibfield  {title} {\enquote {\bibinfo {title} {Robustifying
  twist-and-turn entanglement with interaction-based readout},}\ }\href@noop {}
  {\bibfield  {journal} {\bibinfo  {journal} {arXiv:1803.08789}\ } (\bibinfo
  {year} {2018})}\BibitemShut {NoStop}%
\bibitem [{\citenamefont {Haine}(2018{\natexlab{b}})}]{Haine:2018b}%
  \BibitemOpen
  \bibfield  {author} {\bibinfo {author} {\bibfnamefont {Simon~A.}\
  \bibnamefont {Haine}},\ }\bibfield  {title} {\enquote {\bibinfo {title}
  {Using interaction-based readouts to approach the ultimate limit of detection
  noise robustness for quantum-enhanced metrology in collective spin
  systems},}\ }\href@noop {} {\bibfield  {journal} {\bibinfo  {journal}
  {arXiv:1806.00057}\ } (\bibinfo {year} {2018}{\natexlab{b}})}\BibitemShut
  {NoStop}%
\bibitem [{\citenamefont {Gotlibovych}\ \emph {et~al.}(2014)\citenamefont
  {Gotlibovych}, \citenamefont {Schmidutz}, \citenamefont {Gaunt},
  \citenamefont {Navon}, \citenamefont {Smith},\ and\ \citenamefont
  {Hadzibabic}}]{Gotlibovych:2014}%
  \BibitemOpen
  \bibfield  {author} {\bibinfo {author} {\bibfnamefont {Igor}\ \bibnamefont
  {Gotlibovych}}, \bibinfo {author} {\bibfnamefont {Tobias~F.}\ \bibnamefont
  {Schmidutz}}, \bibinfo {author} {\bibfnamefont {Alexander~L.}\ \bibnamefont
  {Gaunt}}, \bibinfo {author} {\bibfnamefont {Nir}\ \bibnamefont {Navon}},
  \bibinfo {author} {\bibfnamefont {Robert~P.}\ \bibnamefont {Smith}}, \ and\
  \bibinfo {author} {\bibfnamefont {Zoran}\ \bibnamefont {Hadzibabic}},\
  }\bibfield  {title} {\enquote {\bibinfo {title} {Observing properties of an
  interacting homogeneous bose-einstein condensate: Heisenberg-limited momentum
  spread, interaction energy, and free-expansion dynamics},}\ }\href {\doibase
  10.1103/PhysRevA.89.061604} {\bibfield  {journal} {\bibinfo  {journal} {Phys.
  Rev. A}\ }\textbf {\bibinfo {volume} {89}},\ \bibinfo {pages} {061604}
  (\bibinfo {year} {2014})}\BibitemShut {NoStop}%
\bibitem [{\citenamefont {Ketterle}\ \emph {et~al.}(1999)\citenamefont
  {Ketterle}, \citenamefont {Durfee},\ and\ \citenamefont
  {Stamper-Kurn}}]{Ketterle:1999}%
  \BibitemOpen
  \bibfield  {author} {\bibinfo {author} {\bibfnamefont {W.}~\bibnamefont
  {Ketterle}}, \bibinfo {author} {\bibfnamefont {D.~S.}\ \bibnamefont
  {Durfee}}, \ and\ \bibinfo {author} {\bibfnamefont {D.~M.}\ \bibnamefont
  {Stamper-Kurn}},\ }\bibfield  {title} {\enquote {\bibinfo {title} {Making,
  probing and understanding bose-einstein condensates},}\ }in\ \href@noop {}
  {\emph {\bibinfo {booktitle} {Proceedings of the International School of
  Physics ``Enrico Fermi''}}},\ Vol.\ \bibinfo {volume} {140},\ \bibinfo
  {editor} {edited by\ \bibinfo {editor} {\bibfnamefont {M.}~\bibnamefont
  {Inguscio}}, \bibinfo {editor} {\bibfnamefont {S.}~\bibnamefont {Stringari}},
  \ and\ \bibinfo {editor} {\bibfnamefont {C.E.}\ \bibnamefont {Wieman}}}\
  (\bibinfo {year} {1999})\ p.~\bibinfo {pages} {67}\BibitemShut {NoStop}%
\bibitem [{\citenamefont {Hardman}\ \emph
  {et~al.}(2016{\natexlab{b}})\citenamefont {Hardman}, \citenamefont {Wigley},
  \citenamefont {Everitt}, \citenamefont {Manju}, \citenamefont {Kuhn},\ and\
  \citenamefont {Robins}}]{Hardman:2016}%
  \BibitemOpen
  \bibfield  {author} {\bibinfo {author} {\bibfnamefont {K.~S.}\ \bibnamefont
  {Hardman}}, \bibinfo {author} {\bibfnamefont {P.~B.}\ \bibnamefont {Wigley}},
  \bibinfo {author} {\bibfnamefont {P.~J.}\ \bibnamefont {Everitt}}, \bibinfo
  {author} {\bibfnamefont {P.}~\bibnamefont {Manju}}, \bibinfo {author}
  {\bibfnamefont {C.~C.~N.}\ \bibnamefont {Kuhn}}, \ and\ \bibinfo {author}
  {\bibfnamefont {N.~P.}\ \bibnamefont {Robins}},\ }\bibfield  {title}
  {\enquote {\bibinfo {title} {Time-of-flight detection of ultra-cold atoms
  using resonant frequency modulation imaging},}\ }\href {\doibase
  10.1364/OL.41.002505} {\bibfield  {journal} {\bibinfo  {journal} {Opt.
  Lett.}\ }\textbf {\bibinfo {volume} {41}},\ \bibinfo {pages} {2505--2508}
  (\bibinfo {year} {2016}{\natexlab{b}})}\BibitemShut {NoStop}%
\bibitem [{\citenamefont {Stenger}\ \emph {et~al.}(1999)\citenamefont
  {Stenger}, \citenamefont {Inouye}, \citenamefont {Chikkatur}, \citenamefont
  {Stamper-Kurn}, \citenamefont {Pritchard},\ and\ \citenamefont
  {Ketterle}}]{Stenger:1999}%
  \BibitemOpen
  \bibfield  {author} {\bibinfo {author} {\bibfnamefont {J.}~\bibnamefont
  {Stenger}}, \bibinfo {author} {\bibfnamefont {S.}~\bibnamefont {Inouye}},
  \bibinfo {author} {\bibfnamefont {A.~P.}\ \bibnamefont {Chikkatur}}, \bibinfo
  {author} {\bibfnamefont {D.~M.}\ \bibnamefont {Stamper-Kurn}}, \bibinfo
  {author} {\bibfnamefont {D.~E.}\ \bibnamefont {Pritchard}}, \ and\ \bibinfo
  {author} {\bibfnamefont {W.}~\bibnamefont {Ketterle}},\ }\bibfield  {title}
  {\enquote {\bibinfo {title} {Bragg spectroscopy of a bose-einstein
  condensate},}\ }\href {\doibase 10.1103/PhysRevLett.82.4569} {\bibfield
  {journal} {\bibinfo  {journal} {Phys. Rev. Lett.}\ }\textbf {\bibinfo
  {volume} {82}},\ \bibinfo {pages} {4569--4573} (\bibinfo {year}
  {1999})}\BibitemShut {NoStop}%
\bibitem [{\citenamefont {Richard}\ \emph {et~al.}(2003)\citenamefont
  {Richard}, \citenamefont {Gerbier}, \citenamefont {Thywissen}, \citenamefont
  {Hugbart}, \citenamefont {Bouyer},\ and\ \citenamefont
  {Aspect}}]{Richard:2003}%
  \BibitemOpen
  \bibfield  {author} {\bibinfo {author} {\bibfnamefont {S.}~\bibnamefont
  {Richard}}, \bibinfo {author} {\bibfnamefont {F.}~\bibnamefont {Gerbier}},
  \bibinfo {author} {\bibfnamefont {J.~H.}\ \bibnamefont {Thywissen}}, \bibinfo
  {author} {\bibfnamefont {M.}~\bibnamefont {Hugbart}}, \bibinfo {author}
  {\bibfnamefont {P.}~\bibnamefont {Bouyer}}, \ and\ \bibinfo {author}
  {\bibfnamefont {A.}~\bibnamefont {Aspect}},\ }\bibfield  {title} {\enquote
  {\bibinfo {title} {Momentum spectroscopy of 1d phase fluctuations in
  bose-einstein condensates},}\ }\href {\doibase 10.1103/PhysRevLett.91.010405}
  {\bibfield  {journal} {\bibinfo  {journal} {Phys. Rev. Lett.}\ }\textbf
  {\bibinfo {volume} {91}},\ \bibinfo {pages} {010405} (\bibinfo {year}
  {2003})}\BibitemShut {NoStop}%
\bibitem [{\citenamefont {Johnsson}\ \emph {et~al.}(2007)\citenamefont
  {Johnsson}, \citenamefont {Haine}, \citenamefont {Hope}, \citenamefont
  {Robins}, \citenamefont {Figl}, \citenamefont {Jeppesen}, \citenamefont
  {Dugu\'e},\ and\ \citenamefont {Close}}]{Johnsson:2007b}%
  \BibitemOpen
  \bibfield  {author} {\bibinfo {author} {\bibfnamefont {Mattias}\ \bibnamefont
  {Johnsson}}, \bibinfo {author} {\bibfnamefont {Simon}\ \bibnamefont {Haine}},
  \bibinfo {author} {\bibfnamefont {Joseph}\ \bibnamefont {Hope}}, \bibinfo
  {author} {\bibfnamefont {Nick}\ \bibnamefont {Robins}}, \bibinfo {author}
  {\bibfnamefont {Cristina}\ \bibnamefont {Figl}}, \bibinfo {author}
  {\bibfnamefont {Matthew}\ \bibnamefont {Jeppesen}}, \bibinfo {author}
  {\bibfnamefont {Julien}\ \bibnamefont {Dugu\'e}}, \ and\ \bibinfo {author}
  {\bibfnamefont {John}\ \bibnamefont {Close}},\ }\bibfield  {title} {\enquote
  {\bibinfo {title} {Semiclassical limits to the linewidth of an atom laser},}\
  }\href {\doibase 10.1103/PhysRevA.75.043618} {\bibfield  {journal} {\bibinfo
  {journal} {Phys. Rev. A}\ }\textbf {\bibinfo {volume} {75}},\ \bibinfo
  {pages} {043618} (\bibinfo {year} {2007})}\BibitemShut {NoStop}%
\bibitem [{\citenamefont {Szigeti}\ and\ \citenamefont
  {Haine}()}]{Stuart_Simon_in_prep}%
  \BibitemOpen
  \bibfield  {author} {\bibinfo {author} {\bibfnamefont {S.~S.}\ \bibnamefont
  {Szigeti}}\ and\ \bibinfo {author} {\bibfnamefont {S.~A.}\ \bibnamefont
  {Haine}},\ }\href@noop {} {}\bibinfo {howpublished} {in
  preparation.}\BibitemShut {Stop}%
\bibitem [{\citenamefont {Gerlich}\ \emph {et~al.}(2007)\citenamefont
  {Gerlich}, \citenamefont {Hackermuller}, \citenamefont {Hornberger},
  \citenamefont {Stibor}, \citenamefont {Ulbricht}, \citenamefont {Gring},
  \citenamefont {Goldfarb}, \citenamefont {Savas}, \citenamefont {Muri},
  \citenamefont {Mayor},\ and\ \citenamefont {Arndt}}]{Gerlich:2007}%
  \BibitemOpen
  \bibfield  {author} {\bibinfo {author} {\bibfnamefont {Stefan}\ \bibnamefont
  {Gerlich}}, \bibinfo {author} {\bibfnamefont {Lucia}\ \bibnamefont
  {Hackermuller}}, \bibinfo {author} {\bibfnamefont {Klaus}\ \bibnamefont
  {Hornberger}}, \bibinfo {author} {\bibfnamefont {Alexander}\ \bibnamefont
  {Stibor}}, \bibinfo {author} {\bibfnamefont {Hendrik}\ \bibnamefont
  {Ulbricht}}, \bibinfo {author} {\bibfnamefont {Michael}\ \bibnamefont
  {Gring}}, \bibinfo {author} {\bibfnamefont {Fabienne}\ \bibnamefont
  {Goldfarb}}, \bibinfo {author} {\bibfnamefont {Tim}\ \bibnamefont {Savas}},
  \bibinfo {author} {\bibfnamefont {Marcel}\ \bibnamefont {Muri}}, \bibinfo
  {author} {\bibfnamefont {Marcel}\ \bibnamefont {Mayor}}, \ and\ \bibinfo
  {author} {\bibfnamefont {Markus}\ \bibnamefont {Arndt}},\ }\bibfield  {title}
  {\enquote {\bibinfo {title} {A kapitza-dirac-talbot-lau interferometer for
  highly polarizable molecules},}\ }\href {http://dx.doi.org/10.1038/nphys701}
  {\bibfield  {journal} {\bibinfo  {journal} {Nat Phys}\ }\textbf {\bibinfo
  {volume} {3}},\ \bibinfo {pages} {711--715} (\bibinfo {year}
  {2007})}\BibitemShut {NoStop}%
\bibitem [{\citenamefont {Sapiro}\ \emph {et~al.}(2009)\citenamefont {Sapiro},
  \citenamefont {Zhang},\ and\ \citenamefont {Raithel}}]{Sapiro:2009}%
  \BibitemOpen
  \bibfield  {author} {\bibinfo {author} {\bibfnamefont {R.~E.}\ \bibnamefont
  {Sapiro}}, \bibinfo {author} {\bibfnamefont {R.}~\bibnamefont {Zhang}}, \
  and\ \bibinfo {author} {\bibfnamefont {G.}~\bibnamefont {Raithel}},\
  }\bibfield  {title} {\enquote {\bibinfo {title} {Atom interferometry using
  kapitza-dirac scattering in a magnetic trap},}\ }\href {\doibase
  10.1103/PhysRevA.79.043630} {\bibfield  {journal} {\bibinfo  {journal} {Phys.
  Rev. A}\ }\textbf {\bibinfo {volume} {79}},\ \bibinfo {pages} {043630}
  (\bibinfo {year} {2009})}\BibitemShut {NoStop}%
\bibitem [{\citenamefont {Gadway}\ \emph {et~al.}(2009)\citenamefont {Gadway},
  \citenamefont {Pertot}, \citenamefont {Reimann}, \citenamefont {Cohen},\ and\
  \citenamefont {Schneble}}]{Gadway:2009}%
  \BibitemOpen
  \bibfield  {author} {\bibinfo {author} {\bibfnamefont {Bryce}\ \bibnamefont
  {Gadway}}, \bibinfo {author} {\bibfnamefont {Daniel}\ \bibnamefont {Pertot}},
  \bibinfo {author} {\bibfnamefont {Ren\'{e}}\ \bibnamefont {Reimann}},
  \bibinfo {author} {\bibfnamefont {Martin~G.}\ \bibnamefont {Cohen}}, \ and\
  \bibinfo {author} {\bibfnamefont {Dominik}\ \bibnamefont {Schneble}},\
  }\bibfield  {title} {\enquote {\bibinfo {title} {Analysis of kapitza-dirac
  diffraction patterns beyond the raman-nath regime},}\ }\href {\doibase
  10.1364/OE.17.019173} {\bibfield  {journal} {\bibinfo  {journal} {Opt.
  Express}\ }\textbf {\bibinfo {volume} {17}},\ \bibinfo {pages} {19173--19180}
  (\bibinfo {year} {2009})}\BibitemShut {NoStop}%
\bibitem [{\citenamefont {Li}\ \emph {et~al.}(2014)\citenamefont {Li},
  \citenamefont {He},\ and\ \citenamefont {Smerzi}}]{Li:2014}%
  \BibitemOpen
  \bibfield  {author} {\bibinfo {author} {\bibfnamefont {WeiDong}\ \bibnamefont
  {Li}}, \bibinfo {author} {\bibfnamefont {Tianchen}\ \bibnamefont {He}}, \
  and\ \bibinfo {author} {\bibfnamefont {Augusto}\ \bibnamefont {Smerzi}},\
  }\bibfield  {title} {\enquote {\bibinfo {title} {Multimode kapitza-dirac
  interferometry with trapped cold atoms},}\ }\href {\doibase
  10.1103/PhysRevLett.113.023003} {\bibfield  {journal} {\bibinfo  {journal}
  {Phys. Rev. Lett.}\ }\textbf {\bibinfo {volume} {113}},\ \bibinfo {pages}
  {023003} (\bibinfo {year} {2014})}\BibitemShut {NoStop}%
\bibitem [{\citenamefont {He}\ and\ \citenamefont {Niu}(2017)}]{He:2017}%
  \BibitemOpen
  \bibfield  {author} {\bibinfo {author} {\bibfnamefont {Tianchen}\
  \bibnamefont {He}}\ and\ \bibinfo {author} {\bibfnamefont {Pengbin}\
  \bibnamefont {Niu}},\ }\bibfield  {title} {\enquote {\bibinfo {title}
  {Multimode kapitza--dirac interferometer on bose--einstein condensates with
  atomic interactions},}\ }\href {\doibase
  https://doi.org/10.1016/j.physleta.2017.01.049} {\bibfield  {journal}
  {\bibinfo  {journal} {Physics Letters A}\ }\textbf {\bibinfo {volume}
  {381}},\ \bibinfo {pages} {1087 -- 1091} (\bibinfo {year}
  {2017})}\BibitemShut {NoStop}%
\bibitem [{\citenamefont {Fekete}\ \emph {et~al.}(2017)\citenamefont {Fekete},
  \citenamefont {Chai}, \citenamefont {Gardiner},\ and\ \citenamefont
  {Andersen}}]{Fekete:2017}%
  \BibitemOpen
  \bibfield  {author} {\bibinfo {author} {\bibfnamefont {J.}~\bibnamefont
  {Fekete}}, \bibinfo {author} {\bibfnamefont {S.}~\bibnamefont {Chai}},
  \bibinfo {author} {\bibfnamefont {S.~A.}\ \bibnamefont {Gardiner}}, \ and\
  \bibinfo {author} {\bibfnamefont {M.~F.}\ \bibnamefont {Andersen}},\
  }\bibfield  {title} {\enquote {\bibinfo {title} {Resonant transfer of large
  momenta from finite-duration pulse sequences},}\ }\href {\doibase
  10.1103/PhysRevA.95.033601} {\bibfield  {journal} {\bibinfo  {journal} {Phys.
  Rev. A}\ }\textbf {\bibinfo {volume} {95}},\ \bibinfo {pages} {033601}
  (\bibinfo {year} {2017})}\BibitemShut {NoStop}%
\bibitem [{\citenamefont {Guarrera}\ \emph {et~al.}(2017)\citenamefont
  {Guarrera}, \citenamefont {Moore}, \citenamefont {Bunting}, \citenamefont
  {Vanderbruggen},\ and\ \citenamefont {Ovchinnikov}}]{Guarrera:2017}%
  \BibitemOpen
  \bibfield  {author} {\bibinfo {author} {\bibfnamefont {V.}~\bibnamefont
  {Guarrera}}, \bibinfo {author} {\bibfnamefont {R.}~\bibnamefont {Moore}},
  \bibinfo {author} {\bibfnamefont {A.}~\bibnamefont {Bunting}}, \bibinfo
  {author} {\bibfnamefont {T.}~\bibnamefont {Vanderbruggen}}, \ and\ \bibinfo
  {author} {\bibfnamefont {Y.~B.}\ \bibnamefont {Ovchinnikov}},\ }\bibfield
  {title} {\enquote {\bibinfo {title} {Distributed quasi-bragg beam splitter in
  crossed atomic waveguides},}\ }\href {\doibase 10.1038/s41598-017-04710-9}
  {\bibfield  {journal} {\bibinfo  {journal} {Scientific Reports}\ }\textbf
  {\bibinfo {volume} {7}},\ \bibinfo {pages} {4749} (\bibinfo {year}
  {2017})}\BibitemShut {NoStop}%
\bibitem [{\citenamefont {Robins}\ \emph {et~al.}(2006)\citenamefont {Robins},
  \citenamefont {Figl}, \citenamefont {Haine}, \citenamefont {Morrison},
  \citenamefont {Jeppesen}, \citenamefont {Hope},\ and\ \citenamefont
  {Close}}]{Robins:2006}%
  \BibitemOpen
  \bibfield  {author} {\bibinfo {author} {\bibfnamefont {N.~P.}\ \bibnamefont
  {Robins}}, \bibinfo {author} {\bibfnamefont {C.}~\bibnamefont {Figl}},
  \bibinfo {author} {\bibfnamefont {S.~A.}\ \bibnamefont {Haine}}, \bibinfo
  {author} {\bibfnamefont {A.~K.}\ \bibnamefont {Morrison}}, \bibinfo {author}
  {\bibfnamefont {M.}~\bibnamefont {Jeppesen}}, \bibinfo {author}
  {\bibfnamefont {J.~J.}\ \bibnamefont {Hope}}, \ and\ \bibinfo {author}
  {\bibfnamefont {J.~D.}\ \bibnamefont {Close}},\ }\bibfield  {title} {\enquote
  {\bibinfo {title} {Achieving peak brightness in an atom laser},}\ }\href
  {\doibase 10.1103/PhysRevLett.96.140403} {\bibfield  {journal} {\bibinfo
  {journal} {Phys. Rev. Lett.}\ }\textbf {\bibinfo {volume} {96}},\ \bibinfo
  {pages} {140403} (\bibinfo {year} {2006})}\BibitemShut {NoStop}%
\bibitem [{\citenamefont {Johnsson}\ and\ \citenamefont
  {Haine}(2007)}]{Johnsson:2007a}%
  \BibitemOpen
  \bibfield  {author} {\bibinfo {author} {\bibfnamefont {Mattias~T.}\
  \bibnamefont {Johnsson}}\ and\ \bibinfo {author} {\bibfnamefont {Simon~A.}\
  \bibnamefont {Haine}},\ }\bibfield  {title} {\enquote {\bibinfo {title}
  {Generating quadrature squeezing in an atom laser through
  self-interaction},}\ }\href {\doibase 10.1103/PhysRevLett.99.010401}
  {\bibfield  {journal} {\bibinfo  {journal} {Phys. Rev. Lett.}\ }\textbf
  {\bibinfo {volume} {99}},\ \bibinfo {pages} {010401} (\bibinfo {year}
  {2007})}\BibitemShut {NoStop}%
\end{thebibliography}%

\end{document}